\documentclass[fleqn,usenatbib]{mnras}
\usepackage[T1]{fontenc}

\DeclareRobustCommand{\VAN}[3]{#2}
\let\VANthebibliography\thebibliography
\def\thebibliography{\DeclareRobustCommand{\VAN}[3]{##3}\VANthebibliography}

\usepackage{graphicx}	
\usepackage{amsmath}
\usepackage{amssymb,lmodern}
\usepackage{siunitx, xcolor, tikz, upquote, bm, subcaption, cprotect, makecell, upgreek, threeparttable}

\graphicspath{{figs/}}

\definecolor{notecolor}{rgb}{0.8,0,0}

\definecolor{lime}{HTML}{A6CE39}
\DeclareRobustCommand{\orcidicon}{\hspace{-3mm}
	\begin{tikzpicture}
	\draw[lime, fill=lime] (0,0) 
	circle [radius=0.16] 
	node[white] {\hspace{0.1mm}{\fontfamily{qag}\selectfont \tiny ID}};
	\draw[white, fill=white] (-0.07,0.1) 
	circle [radius=0.01];
	\end{tikzpicture}
	\hspace{-5mm}
}

\foreach \x in {A, ..., Z}{\expandafter\xdef\csname orcid\x\endcsname{\noexpand\href{https://orcid.org/\csname orcidauthor\x\endcsname}
		{\noexpand\orcidicon}}
}

\DeclareSIUnit\parsec{pc}
\DeclareSIUnit\jansky{Jy}
\newcommand{\ud}{\mathrm{d}}    

\defcitealias{anstey_21}{A21}

\title[Point sources]{Impact of extragalactic point sources on the low-frequency sky spectrum and cosmic dawn global 21-cm measurements}

\author[Mittal et al.]{Shikhar Mittal$^{1,2}$\,\ \orcidA{}\ \ \thanks{E-mail: shikhar.mittal@tifr.res.in, sm2941@cam.ac.uk}, Girish Kulkarni$^1$\,\ \orcidB{}\ , Dominic Anstey$^{2,3}$\,\ \orcidC{}\ , and Eloy de Lera Acedo$^{2,3}$\,\ \orcidD{}\\
\\
$^1$Tata Institute of Fundamental Research, Homi Bhabha Road, Mumbai 400005, India\\
$^{2}$Battcock Centre for Experimental Astrophysics, Cavendish Laboratory, J.~J.\ Thomson Avenue, Cambridge CB3 0HE, UK\\
$^{3}$Kavli Institute for Cosmology, University of Cambridge, Madingley Road, Cambridge CB3 0HA, UK}

\date{Accepted ---. Received ---; in original form ---}

\pubyear{2024}

\begin{document}
\label{firstpage}
\pagerange{\pageref{firstpage}--\pageref{lastpage}}
\maketitle

\begin{abstract}
Contribution of resolved and unresolved extragalactic point sources to the low-frequency sky spectrum is a potentially non-negligible part of the astrophysical foregrounds for cosmic dawn 21-cm experiments. The clustering of such point sources on the sky, combined with the frequency-dependence of the antenna beam, can also make this contribution chromatic. By combining low-frequency measurements of the luminosity function and the angular correlation function of extragalactic point sources, we develop a model for the contribution of these sources to the low-frequency sky spectrum. Using this model, we find that the contribution of sources with flux density $>10^{-6}\,$Jy to the sky-averaged spectrum is smooth and of the order of a few kelvins at 50--$200\,$MHz. We combine this model with measurements of the galactic foreground spectrum and weigh the resultant sky by the beam directivity of the conical log-spiral antenna planned as part of the Radio Experiment for the Analysis of Cosmic Hydrogen (\textit{REACH}) project. We find that the contribution of point sources to the resultant spectrum is $\sim0.4\%$ of the total foregrounds, but still larger by at least an order of magnitude than the standard predictions for the cosmological 21-cm signal. As a result, not accounting for the point-source contribution leads to a systematic bias in 21-cm signal recovery. We show, however, that in the \textit{REACH} case, this reconstruction bias can be removed by modelling the point-source contribution as a power law with a running spectral index. We make our code publicly available as a Python package labelled \texttt{epspy}.
\end{abstract}


\section{Introduction}

The cosmological 21-cm signal holds a wealth of information about the dark ages, cosmic dawn and epoch of reionisation \citep{Furlanetto_2006, Pritchard_2012, Liu_2020, Shima_2022, Bera_23}. It is well-known, however, that the detection of this signal is extremely challenging due to a number of reasons, such as the rapidly changing ionosphere \citep{Datta_2016, Jordan17, Shen21}, radio frequency interference, noise, and systematic errors \citep{Offringa15, Scheutwinkel_2022, Leeney23, Anstey_23}, chromatic response of the radio antenna \citep{Vedantham, Mozdzen, Wang}, and most importantly the astrophysical foregrounds \citep{pengo_2003, Gnedin_2004, Santos_2005, Bowman_2006, Jelic_2008}. While the strength of the cosmological signal of interest is expected to be of the order of $\sim\SI{100}{\milli\kelvin}$, the astrophysical foregrounds can easily be 10s or 100s of kelvins. These foregrounds are composed of galactic and extragalactic contributions. Primarily, synchrotron radiation due to accelerating relativistic electrons in the galactic magnetic field and free–free emission from the ionized hydrogen gives rise to diffuse galactic emissions \citep{condon,shaver99}, which can be $\sim75$\% of the total foregrounds \citep{Bernardi09} at $\SI{150}{\mega\hertz}$. The remaining part is mostly due to discrete extragalactic radio sources, only some of which are resolved by radio surveys.

In this work we model the contribution of the extragalactic radio point sources to the foregrounds for cosmic dawn \mbox{21-cm} experiments. Such contamination by extragalactic point sources has previously been studied in the context of CMB experiments \citep[][and references therein]{Gonzalez_2005}. We also investigate the effect that point sources have on the detectability of cosmic dawn global 21-cm signal. In this work, we focus on the case of the Radio Experiment for the Analysis of Cosmic Hydrogen (\textit{REACH})\footnote{\url{https://www.astro.phy.cam.ac.uk/research/research-projects/reach}} \citep{Eloy, reach} experiment. Nonetheless, our methods are also applicable to other global 21-cm experiments such as Experiment to Detect the Global EoR Signal \citep[\textit{EDGES},][]{edges}, Shaped Antenna measurement of the background RAdio Spectrum \citep[\textit{SARAS},][]{Patra, saras3}, Large Aperture Experiment to Detect the Dark Ages \citep[\textit{LEDA},][]{Price}, Probing Radio Intensity at high-Z from Marion \citep[\textit{PRIzM},][]{philip}, and Mapper of the IGM Spin Temperature \citep[\textit{MIST},][]{Monsalve_2024}, as well as to experiments targeting the dark ages 21-cm signal by means of lunar-based telescopes such as Dark Ages Radio Explorer \citep[\textit{DARE},][]{Burns_2012, Burns_2017}.

This paper is organized as follows. In section~\ref{sec:method}, we describe our empirical model of extragalactic point sources. In section~\ref{sec:chroma} we investigate the contribution of these sources to the low-frequency sky spectrum as seen by \textit{REACH}. In section~\ref{sec:data} we discuss the implications of point sources contribution on the 21-cm signal extraction. We summarise our conclusions in section~\ref{sec:conc}.


\section{A model for point sources}\label{sec:method}

Our aim is to model the contribution $T_{\mathrm{ps}}$ to the sky spectrum by extragalactic point radio sources as a function of position on the sky, $\bm{\hat{n}}$, and frequency, $\nu$. Thus, we seek $T_{\mathrm{ps}}=T_{\mathrm{ps}}(\bm{\hat{n}},\nu)$. We construct a model for this by incorporating three observational inputs, namely, i) measurements of the flux density distribution, $\ud n/\ud S$, ii) measurements of the clustering of point sources on the sky, $C(\chi)$, and iii) the spectral energy distribution, $S=S(\nu)$, of the sources.

\subsection{Flux density distribution}
Several measurements of the flux density distribution of point sources exist in literature at different survey frequencies and sky coverage. \citet{Baldwin}, \citet{Hales_1988}, and \citet{gilchrist} obtained some of the earliest differential source counts based on radio catalogues by \textit{6C} and \textit{7C} surveys at $\SI{151}{\mega\hertz}$ with the faintest source resolved at $\SI{0.1}{\jansky}$. While some modern surveys have a similar sensitivity to the older survey, a few of the new generation low-frequency telescopes are uncovering radio sky with unprecedented depth and sensitivity. \citet{Intema17} obtained the flux density distribution based on the first alternative data release of the TIFR \textit{GMRT} Sky Survey (\textit{TGSS ADR1}) by the Giant Metrewave Radio Telescope (\textit{GMRT}) covering 90\% of the sky at $\SI{150}{\mega\hertz}$ with the faintest source resolved having flux density of $\SI{0.1}{\jansky}$. \citet{Mandal_2021} did the same based on deep LOFAR Two Meter Sky Survey -- \textit{LoTSS Deep Fields} which cover the entire northern sky at $\SI{150}{\mega\hertz}$ with the faintest source resolved having flux density of $\SI{2.2e-4}{\jansky}$. As per our knowledge \textit{LoTSS Deep Fields} resolves the faintest, $\sim\SI{0.1}{\milli\jansky}$, of the sources till date at $\SI{150}{\mega\hertz}$. \citet{Franzen_2019} did the same based on GaLactic and Extragalactic All-sky MWA (\textit{GLEAM}) at several frequencies between 72 and $\SI{231}{\mega\hertz}$ which covers entire sky south of declination $\SI{+30}{\degree}$. See also the latest work by \citet{Hale_2023} and \citet{tompkins} who have compiled data from a number of surveys at various frequencies.

\begin{figure}
    \centering
    \includegraphics[width=1\linewidth]{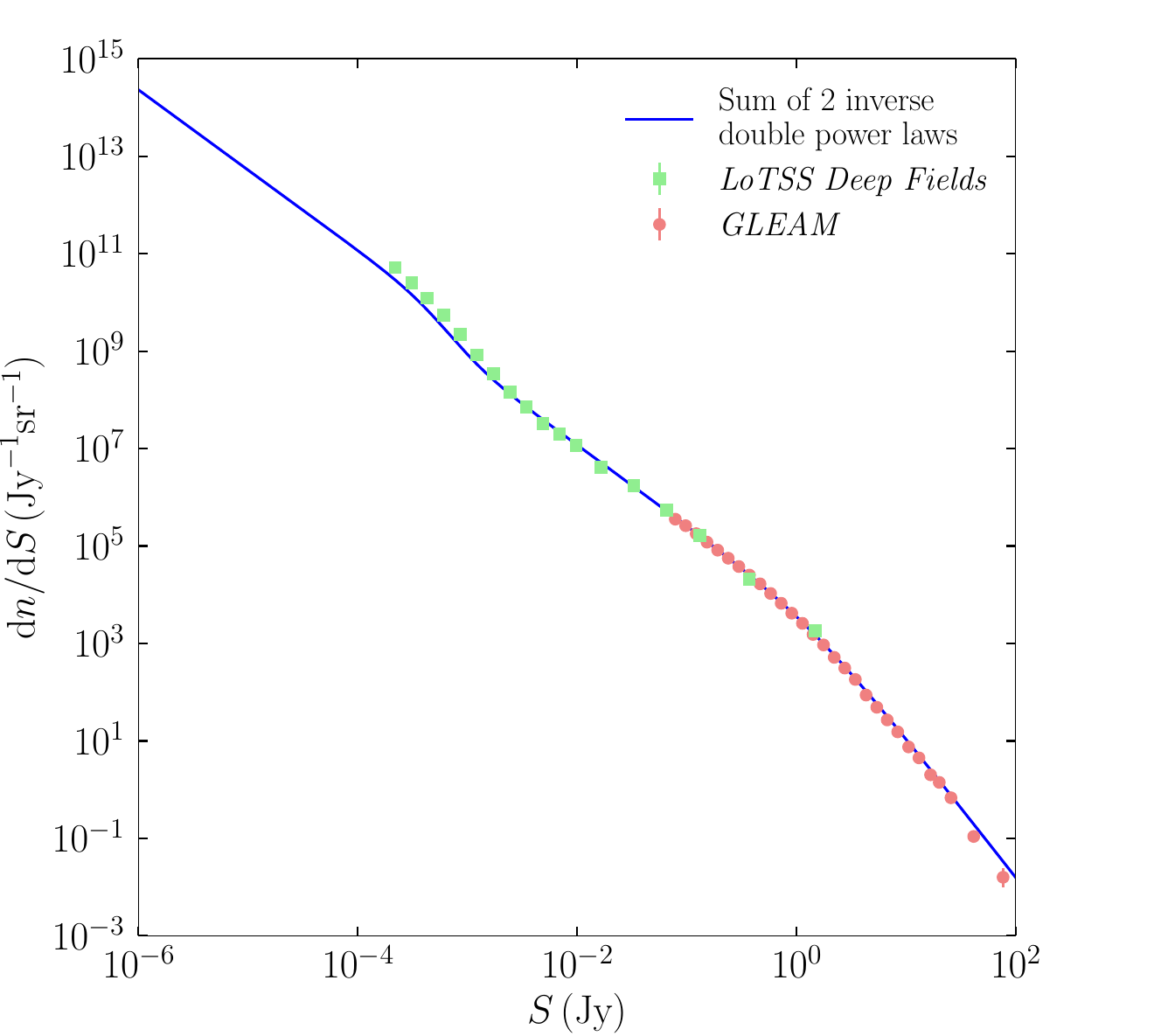}
    \caption{Distribution of flux density of the extragalactic point sources at our chosen reference frequency of $\nu_0=\SI{150}{\mega\hertz}$. The blue curve is a sum of two inverse double-power-laws (see equation~\ref{eq:dndS1}). The green and red points are obtained from \citet{Mandal_2021} and \citet{Franzen_2019}, respectively. The uncertainty bars are too small to be visible in this figure.}\label{fig:dndS}
\end{figure}

In this work we follow the fitting function for $\ud n/\ud S$ reported by \citet{Gervasi_2008} based on 150-MHz \textit{6C} and \textit{7C} surveys
\begin{equation}
    \frac{\ud n}{\ud S}=S^{-2.5}\left[\frac{1}{A_1S^{a_1}+B_1S^{b_1}}+\frac{1}{A_2S^{a_2}+B_2S^{b_2}}\right]\,,\label{eq:dndS1}
\end{equation}
where the 8 numbers, $A_1, B_1, \ldots$, are empirical fitting parameters summarised in Table~\ref{tab:dndS_params} and $S$ and $\ud n/\ud S$ are in units of Jy and $\si{\jansky^{-1}\steradian^{-1}}$, respectively. See blue curve in Fig.~\ref{fig:dndS}. As evident the fitting function \eqref{eq:dndS1} fits reasonably to the newer, \textit{TGSS-ADR1} and \textit{GLEAM} data points, which are shown in light green squares and light red circles, respectively. The uncertainty bars are too small to be visible on the figure.

\begin{table}
    \caption{Value of fitting parameters for the flux density distribution function at reference frequency of $\sim\SI{150}{\mega\hertz}$. Taken from \citet{Gervasi_2008}. We have adopted the weighted average values for $a_1, b_1, a_2, b_2$ and $B_2/B_1$.}\label{tab:dndS_params}
    \centering
    \def\arraystretch{1.1}
    \begin{tabular}{rl}
        \hline
        Parameter & Value                         \\ \hline
        $A_1$     & $(1.65\pm0.02)\times 10^{-4}$ \\
        $B_1$     & $(1.14\pm0.04)\times10^{-4}$  \\
        $a_1$     & $-0.854\pm0.007$              \\
        $b_1$     & $0.37\pm0.02$                 \\
        $a_2$     & $-0.856\pm0.021$              \\
        $b_2$     & $1.47\pm0.15$                 \\
        $A_2/A_1$ & $0.24\pm0.04$                 \\
        $B_2/B_1$ & $(1.8\pm0.2)\times10^7$       \\
        \hline
    \end{tabular}
\end{table}

The fitting function~\eqref{eq:dndS1} implies that the total flux density converges to a finite value as $S$ approaches 0 but the number of sources increase indefinitely. Since we need to scale the flux density for individual sources to a range of frequencies we avoid modelling of infinite number of sources -- without sacrificing the accuracy -- by setting the lower limit of flux density of the point sources to $S_{\mathrm{min}}=10^{-6}\,$Jy \citep{Matteo_2002, Wang_2006, Liu09}.

Previous works investigated impact of only the unresolved point sources. Accordingly, they chose cut-off value above which the point sources can be resolved. For example, \citet{Wang_2006} and \citet{Liu11} adopt $S_{\mathrm{max}}$ to be $\num{e-4}$ and $\SI{0.1}{\jansky}$, respectively. In this work we do not make any distinction between unresolved and resolved point sources, and account for their contribution alike. For our choice of $\ud n/\ud S$ it is extremely rare to find sources above $\SI{e2}{\jansky}$. Stated differently, extending our $S$ range beyond $\SI{e2}{\jansky}$ does not impact the collective foregrounds by more than a few percent \citep{Gervasi_2008}. Hence, we chose $S_{\mathrm{max}}=\SI{e2}{\jansky}$. We do not consider the effect of brighter sources here.

\subsection{Clustering}
At zeroth level one expects the brightness temperature contributed by the point sources to be isotropic with the sources having a random distribution and fluctuation in the number count to be Poissonian. However, a closer look suggests that the anisotropies in the large-scale structure of the Universe will be imprinted on the distribution of sources \citep{Peebles93, Blake_04, Wake_2008}. Consequently, we account for the clustering of point sources.

The anisotropy in the point source positioning, or simply the clustering, can be quantified by the 2-point angular correlation function (2PACF) of the fluctuation of the overdensity $\delta_{\mathrm{ps}}=\delta_{\mathrm{ps}}(\bm{\hat{n}})$, i.e.,
\begin{equation}
    \delta_{\mathrm{ps}}=\frac{n_{\mathrm{ps}}}{\bar{n}_{\mathrm{ps}}}-1\,,
\end{equation}
where $\bar{n}_{\mathrm{ps}}$ is the mean number of point sources per pixel and $n_{\mathrm{ps}}=n_{\mathrm{ps}}(\bm{\hat{n}})$ is the number of point sources per pixel on the clustered sky. Here $\bm{\hat{n}}$ represents the coordinates of a pixel on the sky. Under the assumption of mean isotropy of the Universe, the 2PACF depends only the relative positions represented by $\bm{\hat{n}}'$ and $\bm{\hat{n}}$. Thus,
\begin{equation}
    C(\chi)=\langle \delta_{\mathrm{ps}}(\bm{\hat{n}})\delta_{\mathrm{ps}}(\bm{\hat{n}}+\chi)\rangle\,,
\end{equation}
where $\chi$ is the angle between $\bm{\hat{n}}'$ and $\bm{\hat{n}}$ (given by $\cos\chi=\bm{\hat{n}}'\cdot\bm{\hat{n}}$) \citep{Peebles93, Peacock_2010}.


A number of 2PACF have been derived at different frequencies and for different flux density range. For a summary of existing and ongoing surveys reporting the clustering see \citet{rana18}. See also the latest work by \citet{Hale_2023} who derive 2PACF based on \textit{LoTSS} data at $\SI{144}{\mega\hertz}$. It is evident from the inspection of results that the clustering depends on frequency of observation. Moreover, clustering law is expected to change in different flux density ranges as the brighter sources -- possibly residing in heavier dark matter haloes -- will cluster more strongly than the fainter sources \citep{Cress_1996, Overzier}. As we employ the $\ud n/\ud S$ function at $\SI{150}{\mega\hertz}$ we adopt the 2PACF at the same frequency for consistency. \citet{rana18} derive the 2PACF based on the \textit{TGSS-ADR1} survey at $\nu_0=\SI{150}{\mega\hertz}$. For the threshold flux density of $\SI{50}{\milli\jansky}$ the 2PACF is
\begin{equation}
    C(\chi)=A\chi^{-\gamma}\,,\label{2pacf}
\end{equation}
where $\gamma = 0.821$, $A=7.8\times10^{-3}$ and $\chi$ is in degrees. In this work we assume the same 2PACF is applicable for our choice of flux density range, which goes from $S_{\mathrm{min}}=\SI{1}{\micro\jansky}$ to $S_{\mathrm{max}}=\SI{e2}{\jansky}$. As we show later, the choice of 2PACF has a negligible impact on our sky-averaged temperature. Thus, the assumption of a uniform 2PACF across our choice of flux density range is an excellent approximation.

\subsection{Spectral energy distribution}
Radio experiments will observe the sky for the global signal at a range of frequencies. We thus need to know how the flux density of a source evolves with frequency. The extragalactic radio sources are believed to be active galactic nuclei (AGNs), radio galaxies and relics, star-forming galaxies and haloes that give free-free emission and intergalactic medium \citep{Glesar_2008, Nitu_2021}. Among these, AGNs dominate the total extragalactic foregrounds. Unlike in the case of galactic foregrounds, range of emission mechanisms for different types of extragalactic sources makes their collective flux density a complicated function of frequency. As we will see later, a power law function with a running index explains the collective brightness temperature (even the contribution to antenna temperature for a conical log-spiral and a hexagonal dipole antenna) spectrum sufficiently well. See Fig.~\ref{fig:vary_ps} and Table~\ref{tab:vary_ps}.

At an individual source level, we assume that the spectral energy distribution (SED) for all types of extragalactic sources is very-well fit by a power law for frequencies of our interest, so that $S(\nu)\propto \nu^{-\alpha}$. However, note that besides these standard non-thermal power-law models, there are other SED models in radio band proposed in literature such as the curved power-law model \citep{Callingham_2017}. We investigate such models in a future work.

In terms of brightness temperature, $S(\nu)\propto \nu^{-\alpha}$ translates to $T(\nu)\propto \nu^{-\beta}$, where $\beta=\alpha+2$. Often the radio sources are categorised into flat-spectrum ($\beta<2.5$) and steep-spectrum ($\beta>2.5$). Steep-spectrum are more common at low frequencies -- such as the frequencies of interest in this work -- while flat-spectrum sources are more common at high frequencies, typically $\nu\gtrsim\SI{2}{\giga\hertz}$ \citep{Peocock_1981}. In order to capture the wide spread in $\beta$ values we follow the strategy by \citet{Liu09}; we assume the indices to be normally distributed around $\beta_0$, i.e.,
\begin{equation}
    \mathcal{P}(\beta)=\frac{1}{\sqrt{2\pi}\sigma_{\beta}}\exp\left[-\frac{(\beta-\beta_0)^2}{2\sigma_{\beta}^2}\right]\,,\label{pbeta}
\end{equation}
where we take $\beta_0=2.681$, inferred by \citet{Gervasi_2008} and a spread of $\sigma_{\beta}=0.5$ \citep{Tegmark_2000}. A Gaussian distribution of indices is motivated by \textit{TGSS-ADR1} and \textit{NVSS} surveys data \citep{Intema17, Tiwari_2019}. Our chosen $\beta_0$ value is consistent with steep-spectrum which dominate the point source population at low frequencies.\\

We summarise our choice of free parameters in Table~\ref{Tab:fid}.

\subsection{Modelling point sources}\label{sec:ps}

In this section we put together the 3 observational inputs to simulate the foregrounds contributed by the point sources. There are three main steps to simulate point sources. In brief they are as follows:

\begin{enumerate}
    \item[1)] Find the total number of point sources given an $S$ distribution
    \item[2)] Distribute these sources on the sky based on the clustering law
    \item[3)] Pixel-wise compute the flux density at any frequency and convert it to brightness temperature
\end{enumerate}

We now go over each step in detail. For our first step we find the total number of point sources on the whole sky as
\begin{equation}
    N_{\mathrm{ps}}=\iint_{S_{\mathrm{min}}}^{S_{\mathrm{max}}}\frac{\ud n}{\ud S}\,\ud S\,\ud \Omega=4\pi\int_{S_{\mathrm{min}}}^{S_{\mathrm{max}}}\frac{\ud n}{\ud S}\,\ud S\,.\label{eq:Ns}
\end{equation}
Using equation~\eqref{eq:dndS1} for $\ud n/\ud S$ we get $N_{\mathrm{ps}}\approx 4.4\times10^9$. Thus, we have $4.4\times10^9$ point sources on the whole sky in the flux density range $10^{-6}$ and $\SI{e2}{\jansky}$ at a frequency of $\nu_0=\SI{150}{\mega\hertz}$. The total flux density is given by
\begin{equation}
    S_{\mathrm{tot}}=4\pi\int_{S_{\mathrm{min}}}^{S_{\mathrm{max}}}S\frac{\ud n}{\ud S}\,\ud S\,,
\end{equation}
which comes out to be $\approx\SI{3.26e5}{\jansky}$. As mentioned previously, a lower $S_{\mathrm{min}}$ and a higher $S_{\mathrm{max}}$ than our current choice changes this number by $\sim1\%$.

Next we distribute $N_{\mathrm{ps}}$ sources on the sky given the 2PACF, equation~\eqref{2pacf}. We begin by converting 2PACF to angular power spectrum, $C_\ell$. The following is the conversion relation
\begin{equation}
    C_\ell=2\pi\int_0^{\pi}C(\chi)P_\ell(\cos\chi)\sin\chi\,\mathrm{d}\chi\,,\label{cell}
\end{equation}
where $P_\ell$ are the Legendre polynomials. We use the python package \verb|transformcl|\footnote{\url{https://cltools.readthedocs.io/transformcl/index.html}} \citep{Tessore} for this transformation. Note that the angular power spectrum (APS) obtained in the above equation is dimensionless.

With the APS so obtained we find the corresponding map $\delta_{\mathrm{ps}}$ using the \texttt{synfast} algorithm from \texttt{HEALPix} \citep{Gorski_2005}. We work with \verb|nside| $ =2^9$ so that the number of pixels is $N_{\mathrm{pix}}=3145728$. Finally, the point source distribution in terms of number of sources per pixel is
\begin{equation}
    n_{\mathrm{ps}}(\bm{\hat{n}})=\bar{n}_{\mathrm{ps}}[1+\delta_{\mathrm{ps}}(\bm{\hat{n}})]\,,\label{eq:ncl}
\end{equation}
where $\bar{n}_{\mathrm{ps}}=N_{\mathrm{ps}}/N_{\mathrm{pix}}$, which for our chosen pixelisation and flux density range comes out to be 1402.

\begin{figure}
    \centering
    \begin{subfigure}{0.48\textwidth}
        \includegraphics[width=1\linewidth]{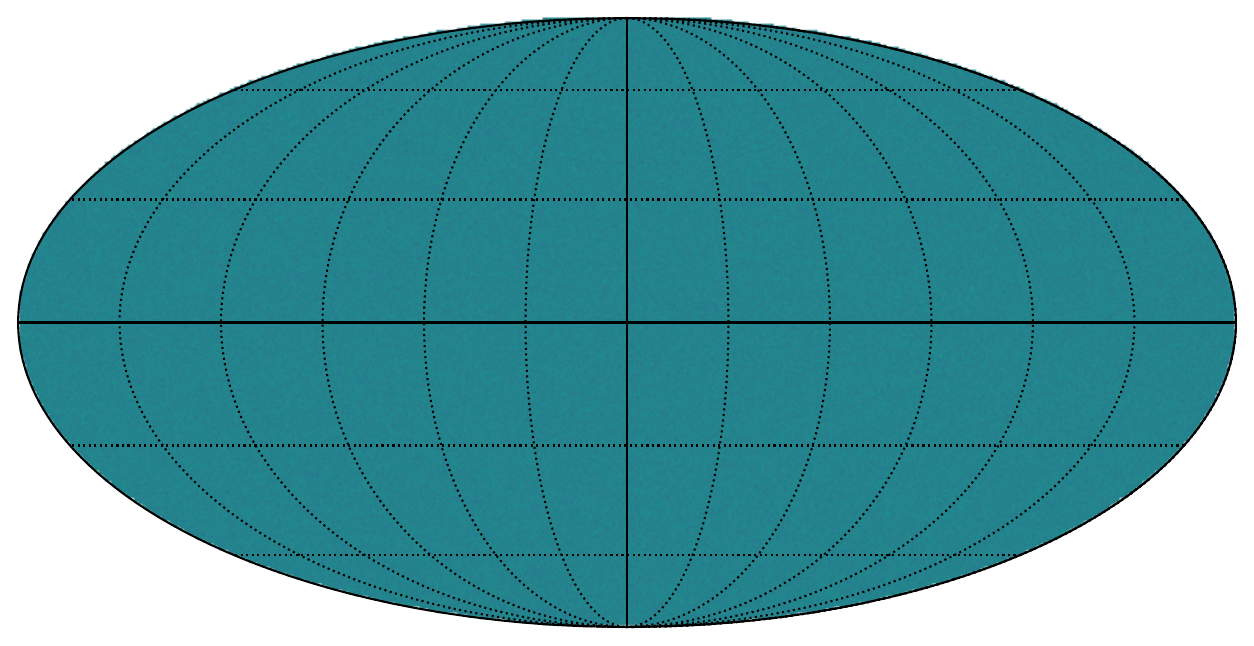}
    \end{subfigure}%
    \\
    \begin{subfigure}{0.48\textwidth}
        \includegraphics[width=1\linewidth]{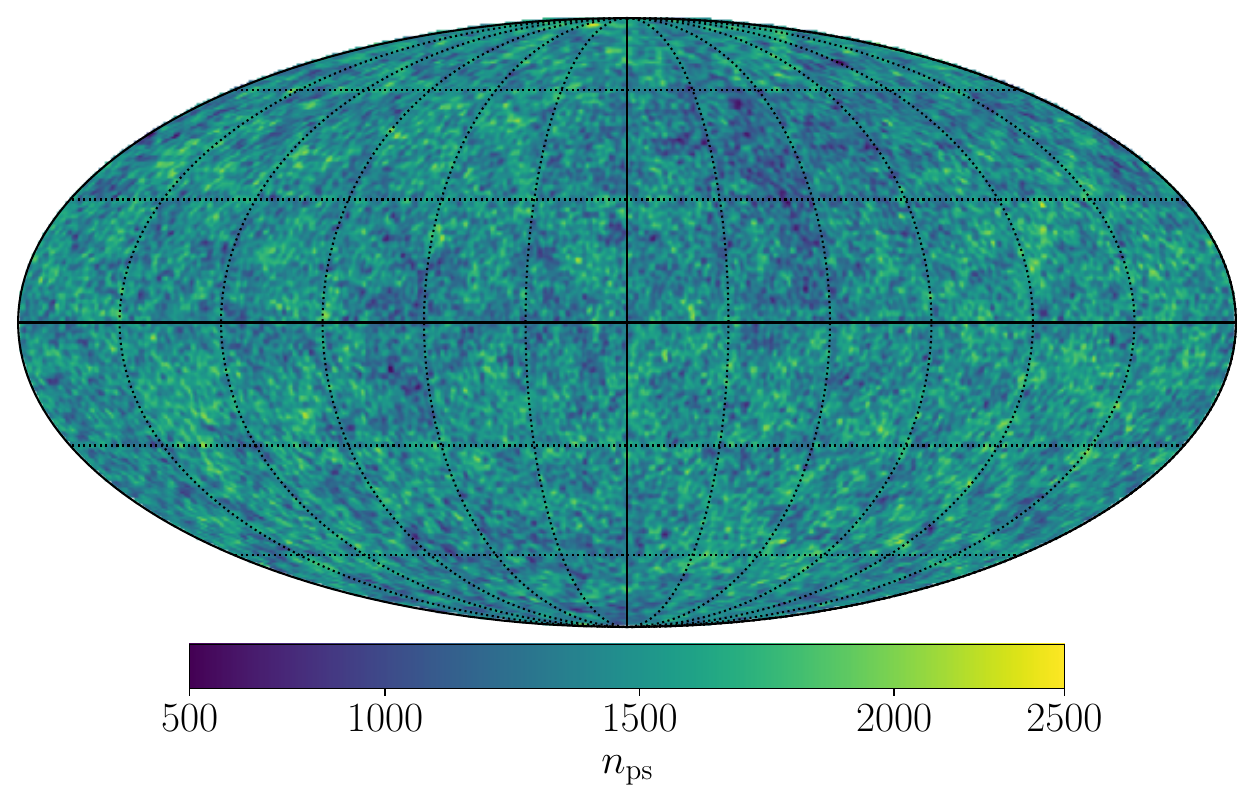}
    \end{subfigure}
    \caption{Top panel shows the number of point sources per pixel for an isotropic sky. In this case the distribution is Poissonian. (Poisson fluctuations are small and hence not visible for the chosen colour range.) Bottom panel shows the number of sources for a clustered sky. The average number of point sources per pixel is 1402 in either case. The colour bar is in linear scale.}\label{fig:sky}
\end{figure}

\begin{figure}
    \centering
    \includegraphics[width=1\linewidth]{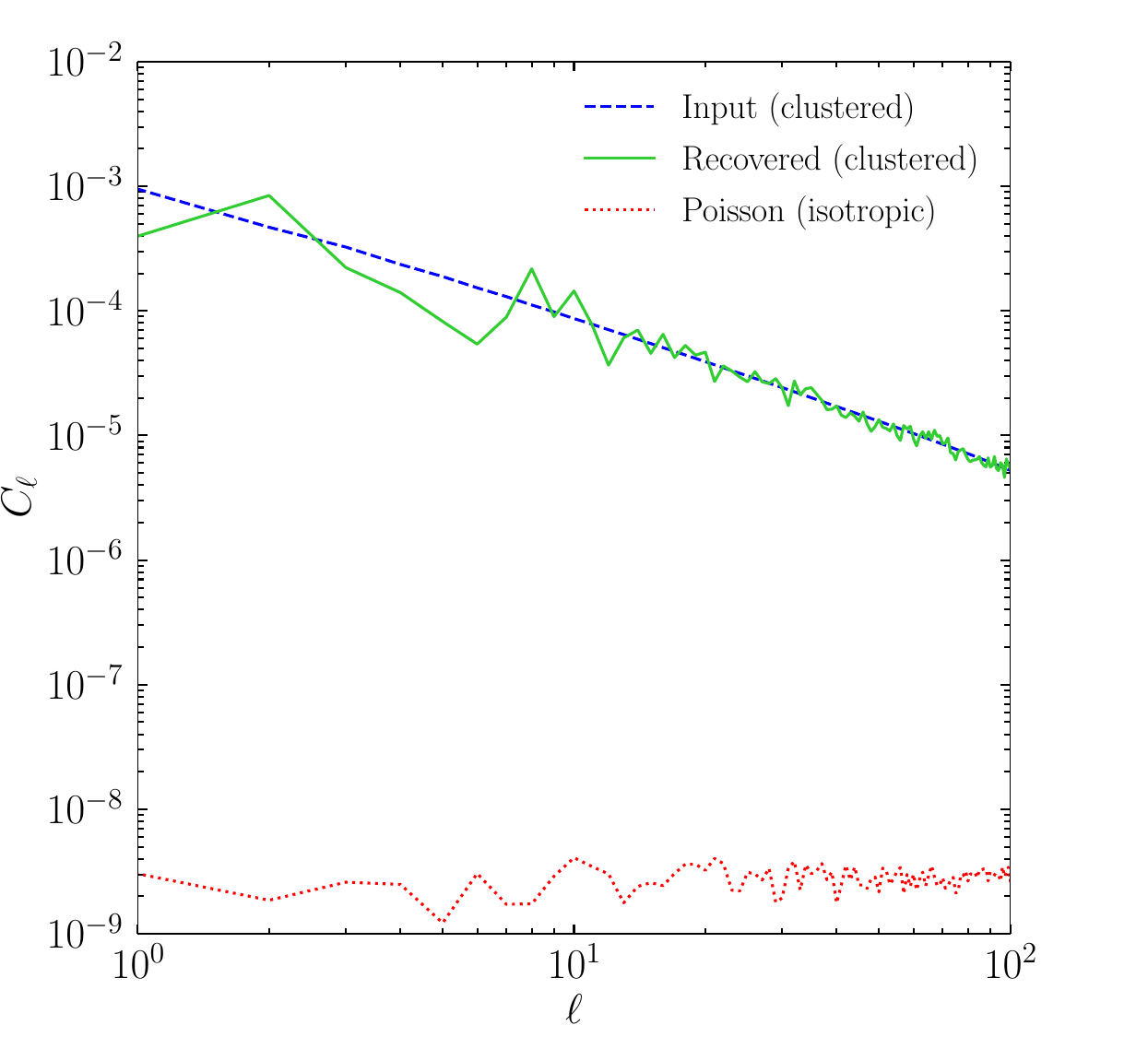}
    \caption{The angular power spectrum of the number density contrast $\delta_{\mathrm{ps}}(\bm{\hat{n}})$. Dotted red curve is for the isotropic sky and blue dashed is for a clustered sky (corresponding to the 2-point angular correlation function in equation~\ref{2pacf}), which also serves as the input for \texttt{synfast}. The green solid curve shows the power spectrum obtained for the simulated map of overdensity, in agreement with the input power spectrum. The corresponding maps of number density are shown in Fig.~\ref{fig:sky}.}\label{fig:aps}
\end{figure}

Figure~\ref{fig:sky} shows the number density (number per pixel) of the extragalactic point sources. The top panel shows an isotropic sky (Poissonian) and the bottom panel shows a clustered sky following the law \eqref{2pacf}\footnote{Given an angular power spectrum there is no unique solution to the fluctuation field. We show in Fig.~\ref{fig:sky} an example realisation from a simulation.}. The contrast between the two skies is immediately apparent by a visual comparison. The sky in the bottom panel shows patchiness which is the result of the clustering of the sources. For a quantitative conclusion one might compare the APS of the overdensity for the two skies. These are shown in Fig.~\ref{fig:aps}. As evident, for an isotropic sky APS is a horizontal line and 0 (albeit with some white noise) but not for the clustered sky.

Having populated our sky, we assign flux density and spectral index to each of the $N_{\mathrm{ps}}$ sources. Let $n_{\mathrm{ps},i}$ be the number of sources on the $i^{\mathrm{th}}$ pixel or a patch on the sky. We draw $n_{\mathrm{ps},i}$ flux densities (they will be at the reference frequency $\nu_0$) from the $S$ distribution (equation~\ref{eq:dndS1}). We convert the flux density to brightness temperature via the standard Rayleigh--Jeans limit of Blackbody function. Putting together everything, the brightness temperature at the $i^{\mathrm{th}}$ pixel due to $j^{\mathrm{th}}$ point source is
\begin{equation}
    T_{\mathrm{ps},ij}(\nu_0)=\frac{(c/\nu_0)^2}{2k_{\mathrm{B}}\Omega_{\mathrm{pix}}}S_{\mathrm{r}}\,,\label{eq1:sed}
\end{equation}
where $k_{\mathrm{B}}$ is the Boltzmann constant, $S_{\mathrm{r}}$ is a random flux density drawn from the distribution and $\Omega_{\mathrm{pix}}=4\pi/N_{\mathrm{pix}}\approx \SI{4e-6}{\steradian}$ (this corresponds to a resolution of $\approx \SI{0.11}{\degree}$) is the solid angle subtended by each pixel.

Next we draw $n_{\mathrm{ps},i}$ spectral indices from the $\beta$ distribution (equation~\ref{pbeta}). Finally, translate the individual brightness temperatures to a frequency $\nu$ (according to the chosen $\beta$) and sum them to get the brightness temperature on the $i^{\mathrm{th}}$ pixel, i.e., $T_{\mathrm{ps},i}(\nu)$,
\begin{equation}
    T_{\mathrm{ps},i}(\nu)=\sum_{j=1}^{n_{\mathrm{ps},i}}T_{\mathrm{ps},ij}(\nu_0)\left(\frac{\nu}{\nu_0}\right)^{-\beta_{ij}}\,.\label{eq2:sed}
\end{equation}
We do the above exercise for all pixels $i=1,2,\ldots N_{\mathrm{pix}}$, and a range of frequencies $\nu$ to obtain the brightness temperature map, $T_{\mathrm{ps}}(\bm{\hat{n}},\nu)$.

\begin{figure*}
    \centering
    \includegraphics[width=0.7\linewidth]{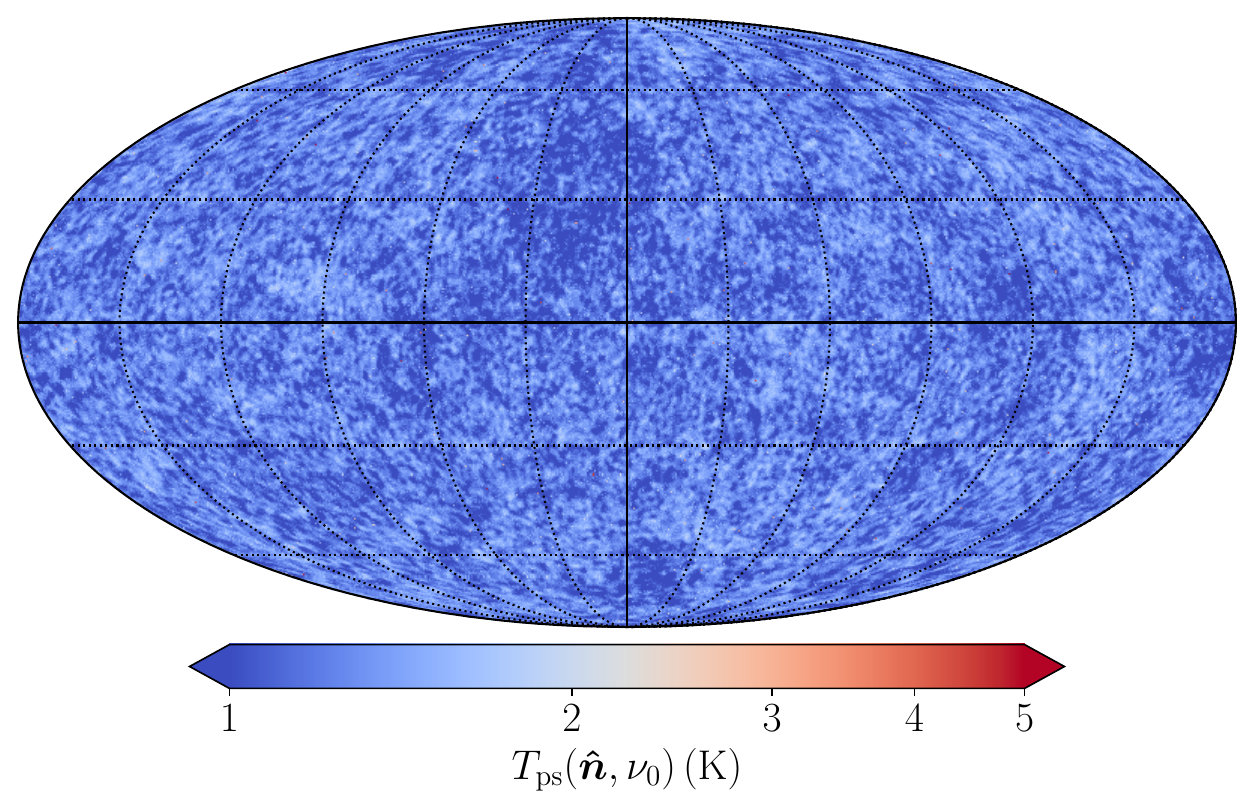}
    \caption{The brightness temperature due to extragalactic point sources at $\nu_0=\SI{150}{\mega\hertz}$. The global or the sky average is $\SI{1.23}{\kelvin}$. The colour bar is in logarithmic scale. The maximum value observed is $\sim\SI{88}{\kelvin}$ but since the majority of pixels have very low brightness, we set the colour bar maximum to $\SI{5}{\kelvin}$ for better visualisation.}\label{fig:Tb}
\end{figure*}

\begin{figure}
    \centering
    \includegraphics[width=1\linewidth]{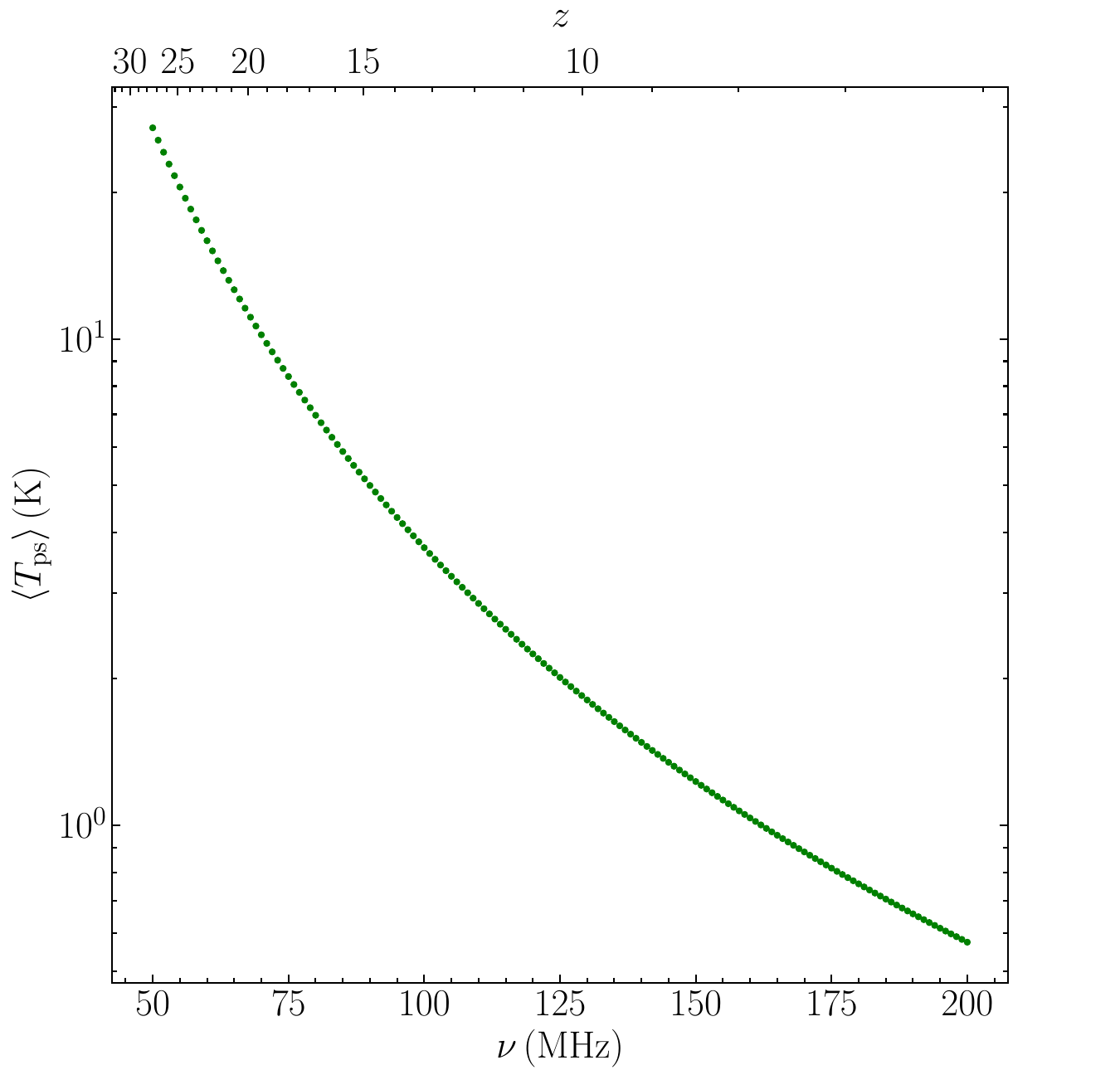}
    \caption{Sky average of the brightness temperature due to the extragalactic point sources in the frequency range 50 to $\SI{200}{\mega\hertz}$.}\label{fig:average_ps}
\end{figure}

Figure~\ref{fig:Tb} shows an example of point sources brightness temperature at a frequency of $\SI{150}{\mega\hertz}$ which corresponds to a redshift of $z=8.4$. For our realisation, the global average is $\SI{1.28}{\kelvin}$. In Fig.~\ref{fig:average_ps} the green-dotted curve shows the global average of the point sources brightness temperature in the frequency range 50 to $\SI{200}{\mega\hertz}$. We represent the global average of the point sources as $\langle T_{\mathrm{ps}}\rangle$ which can be calculated as
\begin{equation}
    \left\langle T_{\mathrm{ps}}\right\rangle (\nu) = \frac{1}{4\pi}\int_{0}^{4\pi} T_{\mathrm{ps}}(\bm{\hat{n}},\nu)\,\ud\Omega\,.
\end{equation}
The curve seen in Fig.~\ref{fig:average_ps} can be described by a power law of index $-2.68$ (additionally with a small running index).

It would be interesting to consider extragalactic point sources as a potential source of the excess radio background (ERB) reported by \textit{ARCADE2}/\textit{LWA1} experiments \citep{Fixsen_2011, Dowell_2018} partly or wholly. (For the latest report on ERB see \citet{Singal_2023}). The spectrum of ERB has a spectral index of $-2.58\pm0.05$, which is close to what point sources imply. This is not a surprising result though, since AGNs dominate our extragalactic source population and it is well-known that the accretion-induced emission from AGNs or AGN-like objects have a strong radio emission which have power law spectrum with index $\approx-2.6$ \citep{Ewall, Ewall2, Mittal_erb}. \citet{Todarello} investigate extragalactic point sources and ERB and concluded that about 20\% of ERB must be of extragalactic origin that traces the large-scale structure.

\begin{table*}
    \centering
    \caption{List of parameters (first column) that control the behaviour of foregrounds due to extragalactic point sources. Second column gives a brief description of these parameters and the third column lists our default choice of values. We do not consider variations in the fitting parameters for the flux density distribution function.}\label{Tab:fid}
    \def\arraystretch{1.1}
    \begin{threeparttable}
    \begin{tabular}{rll}
        \hline
        Parameter          & Description                                & Value               \\ \hline
        $S_{\mathrm{min}}$ & Flux density of the faintest point source  & $\SI{e-6}{\jansky}$ \\
        $S_{\mathrm{max}}$ & Flux density of the brightest point source & $\SI{e2}{\jansky}$  \\
        $\beta_0$          & Mean spectral index of the point sources\tnote{a}   & $2.681$             \\
        $\sigma_\beta$     & Gaussian spread in the spectral indices    & $0.5$               \\
        $A$                & Amplitude of the 2PACF                     & $7.8\times 10^{-3}$ \\
        $\gamma$           & Power-law index of the 2PACF               & $0.821$             \\ \hline
    \end{tabular}
    \begin{tablenotes}
    \item[a] When the SED is expressed in terms of temperature vs frequency, i.e., $T\propto\nu^{-\beta}$.
    \end{tablenotes}
    \end{threeparttable}
\end{table*}

To complete the model for a simulated sky data we account for the galactic emissions. Additionally, for testing our inference pipeline we also inject to this a Gaussian 21-cm signal. Section~\ref{sec:data} gives these details.

Note that for results shown in figures~\ref{fig:Tb} and \ref{fig:average_ps} the effect of chromaticity has not yet been taken into account. This is discussed in the next section.

\section{Contribution to the \textit{REACH} antenna temperature}\label{sec:chroma}

Antenna chromaticity arises because of the dependence of sensitivity of the antenna to the signal on the frequency and direction of observation. As a result of antenna beam chromaticity, the `effective' brightness temperature, or commonly known as the antenna temperature, registered on the antenna in the direction $\bm{\hat{n}}$ and frequency $\nu$ is given by weighting the sky map with directivity pattern of antenna, $D(\bm{\hat{n}},\nu)$, so that $T_{\mathrm{A}}(\bm{\hat{n}},\nu)\sim D(\bm{\hat{n}},\nu)T_{\mathrm{tot}}(\bm{\hat{n}},\nu)$. Thus, the spectrum of sky as seen by the antenna or simply the beam-weighted foregrounds is given by \citep[][hereafter A21]{anstey_21}
\begin{equation}
    T_{\mathrm{A}}(\nu) = \frac{1}{4\pi}\int_0^{4\pi} D(\bm{\hat{n}},\nu)T_{\mathrm{tot}}(\bm{\hat{n}},\nu)\, \ud\Omega + \sigma_{\mathrm{A}}(\nu)\,,\label{eq:chroma1}
\end{equation}
where we also add an uncorrelated Gaussian noise with a standard deviation of $\SI{0.025}{\kelvin}$ representing the antenna noise, $\sigma_{\mathrm{A}}$. Section~\ref{sec:data} gives the details of galactic foregrounds and the total sky temperature, $T_{\mathrm{tot}}=T_{\mathrm{tot}}(\bm{\hat{n}},\nu)$.


For our beam directivity pattern we assume a conical log-spiral antenna as appropriate for \textit{REACH} \citep{Cumner_2022}. We do not work with time varying data because of which we do not have a time integral. Stated differently, we work for a fixed snapshot of the Global Sky Model which we choose to be at UTC 0\,h:\,0\,m:\,0\,s $1^{\mathrm{st}}$ January 2019 at which the Galaxy is above the horizon. The antenna location, height above sea level and orientation (angle between antenna's x-axis and North) are ($\SI{30.71}{\degree}\,$S, $\SI{21.45}{\degree}\,$E), $\SI{1151}{\metre}$ and $\SI{0}{\degree}$, respectively \citep{reach}.

Our focus of investigation in this work will be on the point sources contribution to $T_{\mathrm{A}}$, which may be evaluated as
\begin{equation}
    T_{\mathrm{A,ps}}(\nu) = \frac{1}{4\pi}\int_0^{4\pi} D(\bm{\hat{n}},\nu)T_{\mathrm{ps}}(\bm{\hat{n}},\nu)\,\ud\Omega\,.\label{eq:chroma2}
\end{equation}

\begin{figure}
    \centering
    \includegraphics[width=1\linewidth]{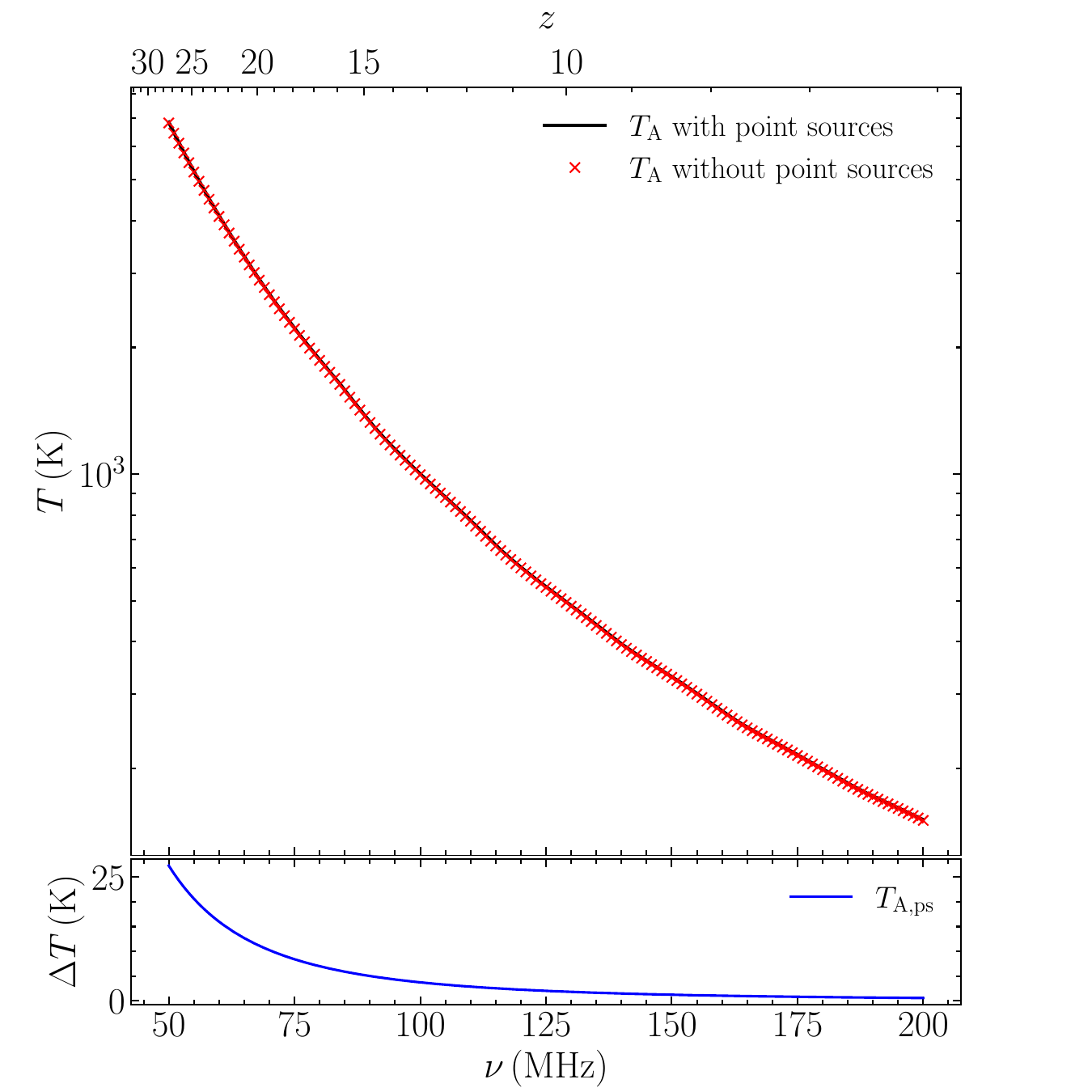}
    \caption{The top panel shows the antenna temperature for two different models of foregrounds. Solid black curve includes and the curve with red crosses does not include the contribution of point sources to the foregrounds. We follow the beam directivity of a conical log-spiral antenna; the \textit{REACH} case. Bottom panel shows the difference between the data represented by solid black and crossed red curves.}\label{fig:spectrum}
\end{figure}

Figure~\ref{fig:spectrum} shows the total antenna temperature for a conical log-spiral antenna given the total sky brightness temperature map, $T_{\mathrm{tot}}(\bm{\hat{n}},\nu)$. Here we have used the fiducial model parameters for point sources given in Table~\ref{Tab:fid}. The red crosses in the top panel show the antenna temperature which does not include the point sources and the solid black curve shows the antenna temperature that includes the point sources emission. The difference between the two, represented as $T_{\mathrm{A,ps}}$, goes from 0.6 to $\SI{27.3}{\kelvin}$. The total antenna temperature (with point sources contribution) goes from 151 to \SI{6851}{\kelvin}. Thus, the point source contribution is $\sim0.4$ percent of the total antenna temperature throughout the frequency range 50 to $\SI{200}{\mega\hertz}$.

\begin{figure}
    \centering
    \includegraphics[width=1\linewidth]{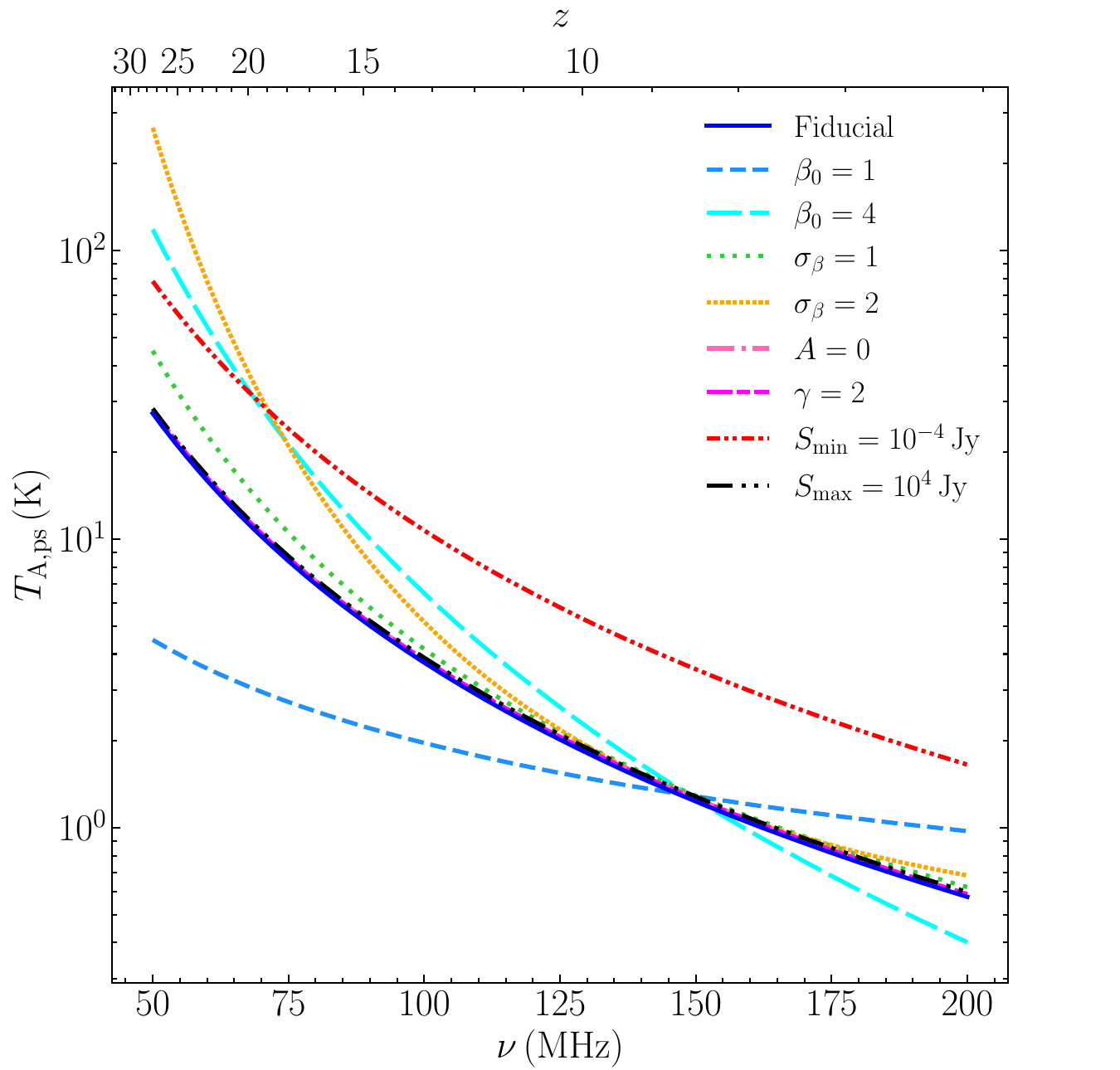}
    \caption{The difference between the antenna temperature with and without point sources contribution. Solid blue is the fiducial model with parameter values given in Table~\ref{Tab:fid} along with the $\ud n/\ud S$ from equation~\eqref{eq:dndS1}. Only one parameter is different as mentioned in the legend while other parameters are maintained at their fiducial value. In all the cases we follow the beam directivity of a conical log-spiral antenna (the \textit{REACH} case).}\label{fig:vary_ps}
\end{figure}

In Fig.~\ref{fig:vary_ps} we show the difference, $T_{\mathrm{A,ps}}$, for various properties of point sources. The thick solid blue curve is repeated from the bottom panel of Fig.~\ref{fig:spectrum}. Models which differ only in $\beta_0$ and $\sigma_\beta$ with respect to the fiducial model will all have the same temperature at the reference frequency ($\SI{150}{\mega\hertz}$ in our case), which is why the blue, light blue, cyan, green and orange curves pass through the same point. This temperature is $\sim\SI{1.3}{\kelvin}$. However, this temperature will scale differently with frequency and hence the shape is different. Allowing sources to have a higher $S$ hardly makes difference. This is because such sources are very rare to find given the $S$ distribution and thus the collective flux remains close to the fiducial value. Increasing $S_{\mathrm{min}}$ to a value of, say, $\SI{e-4}{\jansky}$ though results in a smaller number of sources ($\sim1.6\times10^{8}$), sources with higher $S$ are relatively more likely to be found. This results in stronger extragalactic foregrounds, and hence the antenna temperature, which is why the red curve is above the fiducial model curve. Finally, the variation in clustering law has the least impact on the antenna temperature; pink and magenta curves nearly overlap the fiducial model curve. This is because the different number density distributions we consider for different 2PACFs are only mildly different from each other. As a result we do not see much difference in $T_{\mathrm{ps}}$ maps. Even in the worst case scenario if the beam directivity was a delta function peaked on the pixel of least number of sources (for fiducial model we have $\sim500$; see Fig.~\ref{fig:sky}), number of sources would still be large enough so as to have the same functional form for the temperature as that for the whole sky. Thus, we get nearly the same trend for $T_{\mathrm{A,ps}}$ when we change 2PACF.

We find that $T_{\mathrm{A,ps}}=T_{\mathrm{A,ps}}(\nu)$ can be described by a power-law function with a running spectral index as follows
\begin{equation}
    T_{\mathrm{f}}\left(\frac{\nu}{\nu_0}\right)^{(-\beta_{\mathrm{f}}+\Delta\beta_{\mathrm{f}}\ln{\nu/\nu_0})}\,,\label{eq:psmodel}
\end{equation}
where we fix $\nu_0$ to $\SI{150}{\mega\hertz}$. We fit this form to the different spectra seen in Fig.~\ref{fig:vary_ps}; we find an excellent fit to the data. We have not shown the best-fitting curves as they perfectly overlap, however, Table~\ref{tab:vary_ps} gives the best-fitting parameter values $T_\mathrm{f}, \beta_\mathrm{f}$ and $\Delta\beta_\mathrm{f}$. The 1$\sigma$ uncertainty on all the reported numbers is of the order of $10^{-5}$.
\begin{table}
    \caption{Best-fitting parameters $T_\mathrm{f}, \beta_\mathrm{f}$ and $\Delta\beta_\mathrm{f}$ for power-law-with-a-running-index (equation~\ref{eq:psmodel}) against the extragalactic component of the antenna temperature data using a least-squares fitting. In this table we refer to the various antenna temperature data models by the legend labels shown in Fig.~\ref{fig:vary_ps}. The 1$\sigma$ uncertainty on all the reported numbers is of the order of $10^{-5}$.}\label{tab:vary_ps}
    \centering
    \def\arraystretch{1.1}
    \begin{tabular}{rlll}
        \hline
        Model                      & $T_\mathrm{f}$ & $\beta_\mathrm{f}$ & $\Delta\beta_\mathrm{f}$ \\ \hline
        Fiducial                   & $1.234$        & $2.680$            & $0.126$                  \\
        $\beta_0=1$                & $1.284$        & $1.001$            & $0.124$                  \\
        $\beta_0=4$                & $1.256$        & $4.000$            & $0.126$                  \\
        $\sigma_\beta=1$           & $1.293$        & $2.680$            & $0.501$                  \\
        $\sigma_\beta=2$           & $1.251$        & $2.688$            & $1.996$                  \\
        $A=0$                      & $1.263$        & $2.681$            & $0.125$                  \\
        $\gamma=2$                 & $1.260$        & $2.680$            & $0.126$                  \\
        $S_{\mathrm{min}}=\SI{e-4}{\jansky}$ & $3.540$        & $2.682$            & $0.124$                  \\
        $S_{\mathrm{max}}=\SI{e4}{\jansky}$    & $1.281$        & $2.681$            & $0.125$                  \\
        \hline
    \end{tabular}
\end{table}
From Fig.~\ref{fig:vary_ps} and Table~\ref{tab:vary_ps} it is evident that the flux density range controls the overall amplitude while the spectral index distribution of the radio sources controls the shape of the antenna temperature spectrum. The clustering of sources has a negligible impact.

\begin{figure}
    \centering
    \includegraphics[width=1\linewidth]{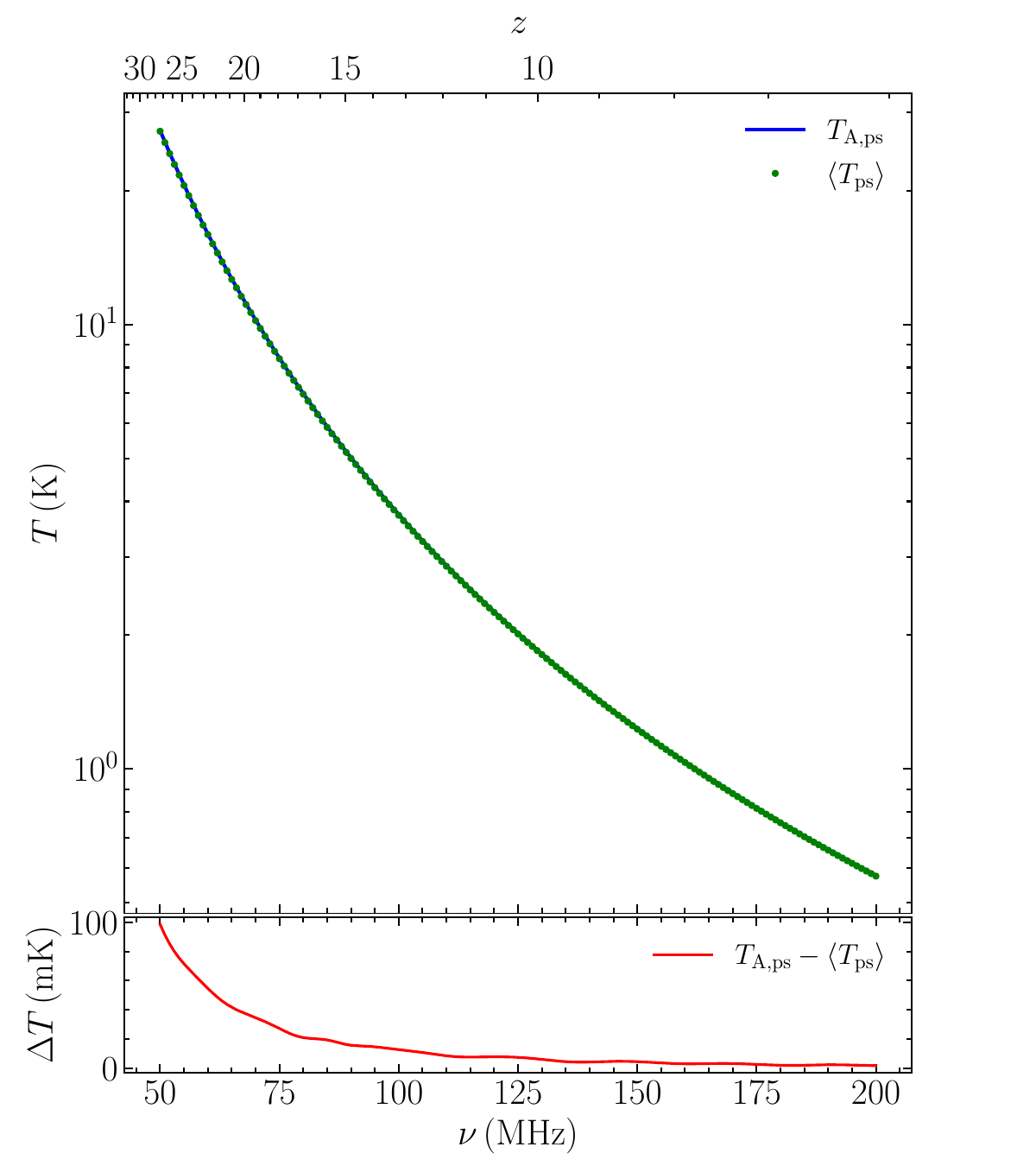}
    \caption{The thick blue solid line is the point sources contribution to antenna temperature (repeated from the bottom panel of Fig.~\ref{fig:spectrum}). Curve with green circles shows the sky average of point sources (repeated from Fig.~\ref{fig:average_ps}). The bottom panel shows the difference between the two curves. We see the chromatic distortions induced by the antenna beam chromaticity.}\label{fig:ps}
\end{figure}

Another interesting result is to compare the point sources spectrum and its contribution to antenna temperature thus explicitly bringing out the chromatic distortions. This is shown in Fig.~\ref{fig:ps} for the fiducial model. The blue solid curve is $T_{\mathrm{A,ps}}$ (which is repeated from the bottom panel of Fig.~\ref{fig:spectrum}) and the green dotted curve (repeated from Fig.~\ref{fig:average_ps}) shows sky average of the point sources, which can also be thought of as $T_{\mathrm{A,ps}}$ for a perfectly achromatic antenna for the full sky, i.e., with $D(\bm{\hat{n}},\nu)=1$ for all $\bm{\hat{n}}$ and $\nu$. The red curve in the bottom panel shows the difference $T_{\mathrm{A,ps}}-\langle T_{\mathrm{ps}}\rangle$. Maximum chromatic distortion is of the order of $\SI{100}{\milli\kelvin}$ seen at $\SI{50}{\mega\hertz}$.

\section{Bias in signal reconstruction due to point sources}\label{sec:data}
So far in this paper we have developed a model for the distribution of extragalactic point sources on the sky. We have also studied the contribution of these sources to the \textit{REACH} beam. We now investigate its effect on the 21-cm signal reconstruction. To do this, we consider a simulated data set which includes the point sources, galactic emissions and a Gaussian 21-cm signal. From this mock dataset, we attempt to extract the 21-cm signal using a Bayesian inference pipeline. We will show that the presence of point sources can bias the 21-cm signal reconstruction.

The strongest component in the low-frequency spectrum measured by a global 21-cm experiment is due to the galactic emissions.  These can be modelled using the Global Sky Model (GSM) maps \citep{costa_2008, zheng_2016}. These maps have been de-sourced for bright extragalactic sources such as giant elliptical galaxies, radio galaxies and quasars. Using GSM, we construct the pure galactic emission map as \citepalias{anstey_21},
\begin{equation}
    T_{\mathrm{gal}}(\bm{\hat{n}},\nu) = \left[T_{\mathrm{gsm}}(\bm{\hat{n}},408)-T_{\mathrm{cmb}}\right]\left(\frac{\nu}{\SI{408}{\mega\hertz}}\right)^{-\beta_{\mathrm{gal}}(\bm{\hat{n}})}\,,
\end{equation}
where $\beta_{\mathrm{gal}}$ is the spectral index for the galactic emissions. The value of $\beta_{\mathrm{gal}}$ on each pixel of the sky can be computed given $T_{\mathrm{gsm}}(\bm{\hat{n}},\nu)$ at two reference frequencies $\SI{230}{\mega\hertz}$ and $\SI{408}{\mega\hertz}$ \citep{costa_2008} so that
\begin{equation}
    \beta_{\mathrm{gal}}(\bm{\hat{n}})=-\frac{\log \frac{T_{\mathrm{gsm}}(\bm{\hat{n}},408)-T_{\mathrm{cmb}}}{T_{\mathrm{gsm}}(\bm{\hat{n}},230)-T_{\mathrm{cmb}}}}{\log \frac{408}{230}}\,.
\end{equation}
We emphasize that galactic emission map $T_{\mathrm{gal}}$ does not include the CMB.

In addition to the foregrounds we have a constant backdrop of CMB, $T_{\mathrm{cmb}}=\SI{2.73}{\kelvin}$. And finally, we add the 21-cm signal. Literature on theoretical modelling of 21-cm signal suggests that at cosmic dawn the signal shape is Gaussian-like \citep{Mirocha_2013, Fialkov14, Mittal_lya, Mittal_pbh}. While there exist theoretical models of the global 21-cm signal in which the absorption feature has a non-Gaussian shape \citep{Mittal_jwst}, here we adopt a Gaussian form given by 
\begin{equation}
    T_{21}(\nu) = -T_{21}^0\exp\left[-\frac{(\nu-\nu_{\mathrm{c}})^2}{2\sigma^2_{21}}\right]\,,
\end{equation}
where we set the amplitude to $T_{21}^0=\SI{0.155}{\kelvin}$, the central frequency to $\nu_{\mathrm{c}}=\SI{85}{\mega\hertz}$ and the Gaussian width to $\sigma_{21}=\SI{15}{\mega\hertz}$.

Weighing the total sky temperature, $T_{\mathrm{tot}}$, by the telescope beam directivity will result in our mock observation of the sky spectrum. (Note that only the galactic and point sources component are spatially varying, while CMB and the 21-cm signal are functions of $\nu$ only.) The goal of global 21-cm signal reconstruction is to extract the 21-cm signal from the observed sky spectrum. 

Given the antenna temperature data we attempt to extract the 21-cm signal. For our inference we follow a fully Bayesian analysis pipeline. In Bayesian statistics, a model $\mathcal{M}$ characterised by parameter set $\theta$ is fit to a data set $\mathcal{D}$ for a given choice of priors. Using Bayes' theorem the posterior distribution, or the probability of getting a certain parameter set can be computed as
\begin{equation}
    \mathrm{P}(\theta|\mathcal{D},\mathcal{M})=\frac{\mathrm{P}(\mathcal{D}|\theta,\mathcal{M})\mathrm{P}(\theta|\mathcal{M})}{\mathrm{P}(\mathcal{D}|\mathcal{M})}\,,
\end{equation}
where $\mathrm{P}(\mathcal{D}|\theta,\mathcal{M})=\mathcal{L}$ is the likelihood, $\mathrm{P}(\theta|\mathcal{M})=\uppi$ is the prior distribution and $\mathrm{P}(\mathcal{D}|\mathcal{M})=\mathcal{Z}$ is the Bayesian evidence. We work with a Gaussian likelihood and write
\begin{equation}
    \ln \mathcal{L}=\sum -\frac{1}{2}\ln(2\pi\sigma'^2)-\frac{1}{2}\left(\frac{T_{\mathcal{D}}-T_{\mathcal{M}}}{\sigma'}\right)^2\,,
\end{equation}
where $\sigma'$ is a uniform noise value across the entire frequency band \citep{Scheutwinkel_2023}. The sum runs over 151 values corresponding to frequencies ranging from 50 to $\SI{200}{\mega\hertz}$ on interval of $\SI{1}{\mega\hertz}$. Note that $\sigma'$ is a free parameter and is therefore part of $\theta$. We implement our Bayesian pipeline using the python package \verb|PolyChord|\footnote{\url{https://github.com/PolyChord/PolyChordLite}} \citep{handley1, handley2}, using the code's default settings throughout. 

\subsection{Reconstruction in the absence of point sources}\label{sec:pipe1}
We first consider the scenario when there is no point-source contribution to the sky. In this case we have
\begin{equation}
    T_{\mathrm{tot}}(\bm{\hat{n}},\nu)=T_{\mathrm{gal}}(\bm{\hat{n}},\nu)+T_{\mathrm{cmb}}+T_{21}(\nu)\,.
\end{equation}
Weighing $T_{\mathrm{tot}}$ by the beam directivity using equation~\eqref{eq:chroma1} gives us the mock data, $T_{\mathcal{D}}$.


For our Bayesian inference pipeline we require an antenna temperature model, $T_{\mathcal{M}}$. \citetalias{anstey_21} showed that non-smooth distortions introduced by chromatic distortions cannot be captured by smooth polynomial foreground fits. Thus, in this work we work with the parametrized sky model, in the same spirit of simulated antenna temperature generation, for the inference procedure. We construct a fitting model for the galactic foregrounds map similar to the construction used for the simulated data; we scale the brightness temperature at a known reference or a base frequency to the required frequency using the spectral index map. These spectral indices will be the free parameters which need to be inferred. However, for a faster likelihood inference while maintaining accuracy, we work with a course-grained version of spectral index map with effectively only $N_{\mathrm{reg}}$ regions or pixels on the sky with different $\beta$'s. Throughout we work with $N_{\mathrm{reg}}=9$. Thus, due to galactic component we introduce 9 free parameters. We hereafter refer to the course-grained galactic foregrounds model by \citetalias{anstey_21} as the `$N$-region' model.

Just as in the case of simulated data construction, we next add the constant CMB piece to the model. This does not have any associated free parameter.

Finally, we assume a Gaussian 21-cm signal characterised by 3 parameters, namely, amplitude, width and the central frequency. Thus, we have our first model to fit to the data
\begin{multline*}
    T_{\mathcal{M}}=(\text{beam directivity})\times(N\text{-region galactic} + \text{CMB} + \\\text{Gaussian 21-cm signal})\,,
\end{multline*}
where, as in the case of simulated data generation, we have weighed the net temperature by the perfectly-known beam directivity. (It might be possible to consider the case of a parametrized beam directivity but we do not explore it in this work.)

Thus, for the inference we have $9$ parameters for the foreground model, 3 parameters for the Gaussian 21-cm model and 1 additional parameter which is the uncorrelated antenna noise, making a total of 13 parameters. For all parameters we use uniform priors.

\begin{figure}
    \centering
    \includegraphics[width=1\linewidth]{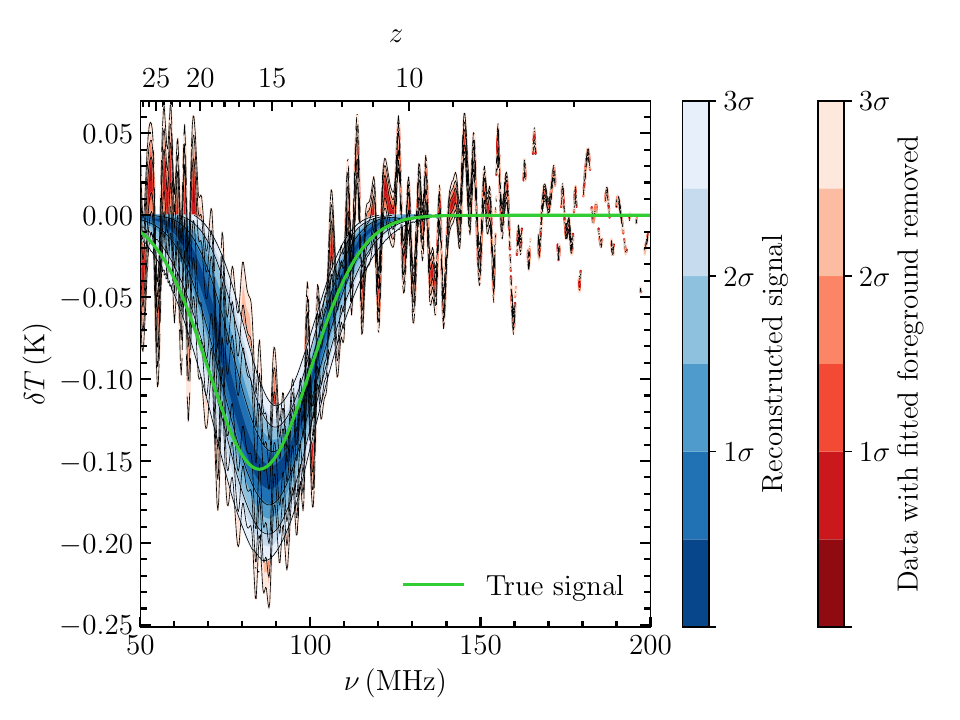}
    \caption{The red region shows the posterior with different sigma levels of the residual after removing a foregrounds model (only galactic) from the mock antenna temperature data. The blue region shows the the posterior with different sigma levels of the best-fitting 21-cm signal. Finally, the green curve shows the true 21-cm signal that was injected to the mock data. The mock data does not contain the extragalactic point sources contribution. The fitting model has a galactic emissions term, CMB and 21-cm signal. The signal recovery is excellent in this case.}\label{fig:pipe1}
\end{figure}

The scenario considered here is a reproduction of the results from \citetalias{anstey_21} (which is the default \textit{REACH} pipeline). In this case we only have galactic emission as part of the foregrounds. Consequently, for inference procedure a course-grained $N$-region foregrounds model suffices as already established by \citetalias{anstey_21}. Figure~\ref{fig:pipe1} shows the recovered or the best-fitting 21-cm signal and its posterior in blue, the residuals after subtraction of $N$-region model of foregrounds in red and the true injected signal in green. As evident the recovered signal is in excellent agreement with true injected signal.

The recovered signal parameters with 1-sigma uncertainty limits are $\nu_{\mathrm{c}} = 88.50 \pm 1.18\,\mathrm{MHz}, \sigma_{21} = 12.43 \pm 1.16\,\mathrm{MHz}$ and $T_{21}^0=0.162\pm 0.0157\,\mathrm{K}$. Thus, the true injected values are within 3-sigma of the inferred values.

\subsection{Mock data with point sources}
We now consider the scenario when we have a finite point sources contribution to the total sky brightness. In this case the total sky data is given by
\begin{equation}
    T_{\mathrm{tot}}(\bm{\hat{n}},\nu)=T_{\mathrm{gal}}(\bm{\hat{n}},\nu)+T_{\mathrm{ps}}(\bm{\hat{n}},\nu)+T_{\mathrm{cmb}}+T_{21}(\nu)\,.
\end{equation}
As before, weighing $T_{\mathrm{tot}}$ by the beam directivity, $D$, gives us the mock antenna temperature data, $T_{\mathcal{D}}$. The point sources contribution that we consider above follows the fiducial set of parameters given in Table~\ref{Tab:fid}. Note that this contribution to antenna temperature data can be described by the power-law function with a running spectral index with parameter values $T = \SI{1.234}{\kelvin}, \beta = 2.680$ and $\Delta\beta = 0.126$ (see `Fiducial' row in Table~\ref{tab:vary_ps}). Just as in the case of 21-cm signal, these values serve as the true `injected' parameter values allowing for a pipeline testing, as we demonstrate in subsection~\ref{sec:improved-piepeline}.

The $T_{\mathrm{tot}}$ constructed above contains galactic as well as extragalactic emission, CMB and 21-cm signal. The brightness temperature data so obtained is a result of a variety of astrophysical and cosmological processes. It is a simple yet sufficiently realistic model representative of the real picture considered for the first time.

For fitting procedure when we have point sources contribution to the data, we consider two sub-cases as described below.
\subsubsection{Signal recovery without correcting for point sources}
For our first sub-case we use exactly the same fitting model as in the previous section so that our model is
\begin{multline*}
    T_{\mathcal{M}}=(\text{beam directivity})\times(N\text{-region galactic} + \text{CMB} + \\\text{Gaussian 21-cm signal})\,.
\end{multline*}
Thus, we have 13 free parameters and we set the same uniform prior ranges as in the previous case.

\begin{figure}
    \centering
    \includegraphics[width=1\linewidth]{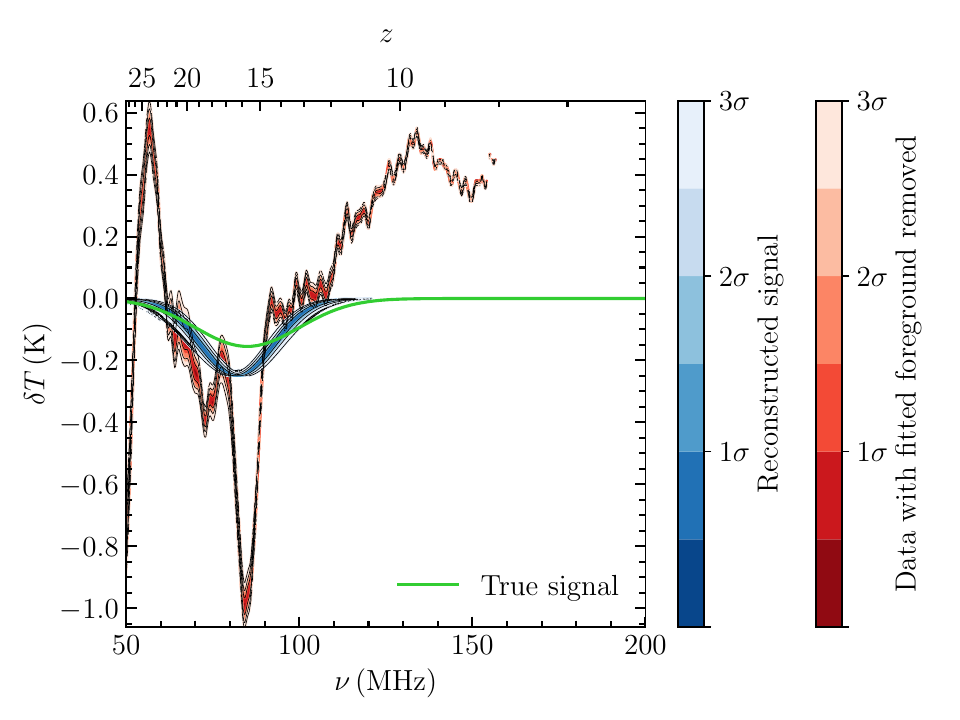}
    \caption{Contrary to the scenario shown in Fig.~\ref{fig:pipe1}, the mock data contains the extragalactic point sources contribution. However, the fitting model remains the same having a galactic term, CMB and 21-cm signal. As evident we do not recover the 21-cm signal.}\label{fig:pipe2}
\end{figure}

As evident from Fig.~\ref{fig:pipe2}, the signal recovery is quite poor in this case. While the data has four components, namely a galactic, an extragalactic, CMB and a 21-cm signal, fitting model used in this pipeline has all but the extragalactic part. Thus, there is no piece in the fitting model to account for the extragalactic foregrounds and as a result the residuals after removing best-fitting foregrounds from simulated data are quite large, as shown by red posteriors. Also, note that value of noise parameter inferred is $\sigma'=(0.1\pm\num{4.7e-5})\,$K which is of the order of the 21-cm signal strength.

The inferred values for the 21-cm signal are $\nu_{\mathrm{c}} = 82.09 \pm 0.81\,\mathrm{MHz}, \sigma_{21} = 10.50 \pm 0.42\,\mathrm{MHz}$ and $T_{21}^0=0.248\pm 0.002\,\mathrm{K}$, which are more than 3-sigma away from the true value, especially the signal depth $T_{21}^0$.

\subsubsection{Improved reconstruction of signal with correction for point sources}\label{sec:improved-piepeline}
For the final case, in our modelling we account for the extragalactic point sources contribution in the data. As we have demonstrated via Fig.~\ref{fig:vary_ps} and Table~\ref{tab:vary_ps}, the point sources contribution to the antenna temperature is a smooth function of frequency -- a power-law-with-a-running-index function. Motivated by this we propose equation~\eqref{eq:psmodel} to account for the point sources contribution present in the antenna temperature data. Thus,
\begin{multline*}
    T_{\mathcal{M}}=(\text{beam directivity})\times(N\text{-region galactic} + \\\text{power-law-with-running-index extragalactic}+\text{CMB} + \\\text{Gaussian 21-cm signal})\,.
\end{multline*}
Correspondingly, we have three additional free parameters $T_{\mathrm{f}}$, $\beta_{\mathrm{f}}$ and $\Delta\beta_{\mathrm{f}}$. So now we have $9+3=12$ parameters for the foreground model, 3 parameters for the Gaussian 21-cm model and 1 additional parameter which is the uncorrelated antenna noise, making a total of 16 parameters. For all parameters we use uniform priors, either in linear scale or log scale. We have a log uniform prior $[10^{-3},10^2]\,$K for $T_{\mathrm{f}}$, uniform prior $[0,5]$ for $\beta_{\mathrm{f}}$ and uniform prior $[0,5]$ for $\Delta\beta_{\mathrm{f}}$.

Note that, as opposed to spectral inhomogeneity in the $N$-region model to capture the galactic emissions in the data, there is no spectral inhomogeneity associated with the point sources model. Simply stated there is no $\bm{\hat{n}}$ dependence.

\begin{figure}
    \centering
    \includegraphics[width=1\linewidth]{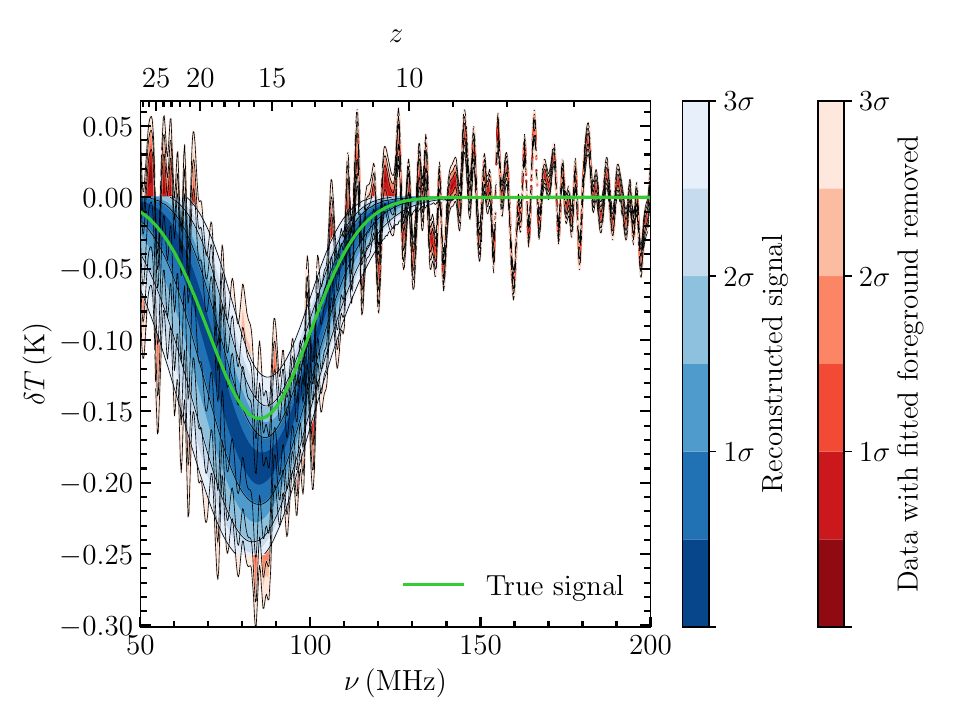}
    \caption{Similar to the case shown in Fig.~\ref{fig:pipe2}, we have an extragalactic contribution to the data. However, the fitting model has an extragalactic foregrounds model in addition to the galactic term, CMB and 21-cm signal. As evident we find that the signal recovery is excellent compared to the scenario shown in Fig.~\ref{fig:pipe2}.}\label{fig:mod3}
\end{figure}

We show our results for this setup in Fig.~\ref{fig:mod3}. As evident the result is quite promising showing that the net sky model, which includes point sources contribution, can be fit reliably using a course-grained $N$-region in conjunction with a 3-parameter power-law-with-a-running-index foregrounds model. The inferred values of point source model parameters are $T_\mathrm{f}=1.072\pm0.041\,\mathrm{K}, \beta_{\mathrm{f}}=2.333\pm 0.102$ and $\Delta\beta_\mathrm{f}=0.067\pm0.045$. Thus, the true `injected' values of the point sources model are within 3-sigma of the inferred values. (Recall that the true `injected' values are $T = \SI{1.234}{\kelvin}, \beta = 2.680$ and $\Delta\beta = 0.126$). Also, the true 21-cm signal parameters are within 1-sigma of the inferred values, $\nu_{\mathrm{c}} = 85.38 \pm 1.61\,\mathrm{MHz}, \sigma_{21} = 14.47 \pm 1.68\,\mathrm{MHz}$ and $T_{21}^0=0.192\pm 0.023\,\mathrm{K}$.

The function power-law-with-a-running-index (equation~\ref{eq:psmodel}) that we introduce in this work, when expressed in log-log units takes the familiar `polynomial' (in $\log T$-$\log\nu$ space) form (precisely a second degree polynomial). Global 21-cm experiments such as \textit{EDGES} \citep{Bowman_2018} and \textit{SARAS} \citep{saras3} have used a polynomial function to model the total foregrounds. However, \citet{anstey_21} have shown that using a polynomial, in general, for the foregrounds (only galactic) can mask the 21-cm signal completely. Nevertheless, in this work it appears that extragalactic point sources contribution to foregrounds admits a polynomial description leading to a reliable inference of cosmological global 21-cm signal, as evidenced by our Fig.~\ref{fig:mod3}. We emphasize that we use a polynomial only to model the extragalactic foreground component. For galactic component we have used $N$-region model throughout this work. It is this fundamental difference which breaks the degeneracy between the galactic and extragalactic component and allows for an excellent signal recovery.\\


It remains to be investigated how the additional presence of ionosphere will affect our extraction of the signal \citep{Shen_22}. Other important effects to account for are the horizon \citep{pattinson}, time-varying systematics \citep{Kirkham}, sinusoidal-like instrumental systematics \citep{Scheutwinkel_2022} and amplitude errors in radio maps \citep{Pagano_2023}. We leave the study of impact of these effects on the signal extraction in the presence of extragalactic point source emission for a future work.


\section{Conclusions}\label{sec:conc}

In this work, we built a model for the contribution of extragalactic point sources to the low-frequency sky spectrum. Using measurements of the luminosity function and the angular correlation function of point sources, our model can compute a full sky with clustered point sources. For a flux limit of $S_\mathrm{min}=10^{-6}\,$Jy, we find the all-sky spectrum has approximately a power-law distribution with typical values of a few kelvins at 50--$200\,$MHz.

Further, we combined our point source model with a model of the galactic foregrounds, CMB and 21-cm signal to simulate the total sky temperature. We then weighed this by the beam pattern of a conical log-spiral antenna to simulate the antenna temperature as measured by the \textit{REACH} experiment. We find that, at a representative frequency of $\SI{85}{\mega\hertz}$ which corresponds to a redshift of $z=15.7$, contribution to the antenna temperature by the extragalactic point sources is $\SI{5.9}{\kelvin}$ while the total antenna temperature is $\SI{1584}{\kelvin}$. Thus, extragalactic contribution is less than a percent compared to the total sky temperature data, as it is throughout the frequency range of interest, 50 to $\SI{200}{\mega\hertz}$. While it seems insignificant compared to the total brightness, extragalactic contribution is still more than an order of magnitude stronger than the standard cosmological 21-cm signal.

We also find that the antenna temperature spectrum corresponding to a conical log-spiral antenna (\textit{REACH} beam) is quite smooth for a wide range of properties of the point sources. Nevertheless, the chromatic distortions induced can be as high as $\SI{0.1}{\kelvin}$, which is of the order of 21-cm signal strength. For any of these models we noted that a power law function with a running spectral index provides an excellent fit, with uncertainty levels of the order of $10^{-5}$, to the point sources spectrum. If unaccounted, in the presence of point sources in the data, the 21-cm signal recovery suffers with severe systematic bias. This bias can be successfully removed by incorporating our point-source model in the reconstruction procedure. This work paves the way forward for a more accurate inference for the current and upcoming global and interferometric 21-cm experiments.

\section*{Acknowledgements}
We thank Harry Bevins, Paul Scott, Saurabh Singh and Nicolas Tessore for comments. It is also a pleasure to acknowledge discussions with several other members of the \textit{REACH} (Radio Experiment for the Analysis of Cosmic Hydrogen) collaboration. GK gratefully acknowledges support by the Max Planck Society via a partner group grant. GK is also partly supported by the Department of Atomic Energy (Government of India) research project with Project Identification Number RTI 4002. DA and EdLA are supported by the Science and Technologies Facilities Council.

\section*{Data Availability}
We release our code for building models of point-source sky as a Python package called \verb|epspy|, available from \verb|PyPI| and GitHub (\url{https://github.com/shikharmittal04/epspy.git}). 

\bibliographystyle{mnras}
\bibliography{biblio}

\appendix
\section{Antenna temperature for a hexagonal dipole antenna}
The currently deployed antenna for the \textit{REACH} experiment is the hexagonal dipole antenna (HDA). \citet{Anstey_22} showed that conical log-spiral antenna is far superior at recovering the 21-cm signal than HDA for various foreground and signal models because HDA is much more chromatic than a conical log-spiral. (However, conical log-spiral is costlier and harder to build than HDA). Phase I of \textit{REACH} experiment operates an HDA. It will, thus, be interesting to consider the antenna temperature as observed through an HDA.

In this appendix we present our results shown in figures \ref{fig:spectrum} to \ref{fig:ps} for an HDA. For this antenna we have beam directivity data for a frequency range of 50 - $\SI{150}{\mega\hertz}$. 

\begin{figure}
    \centering
    \includegraphics[width=1\linewidth]{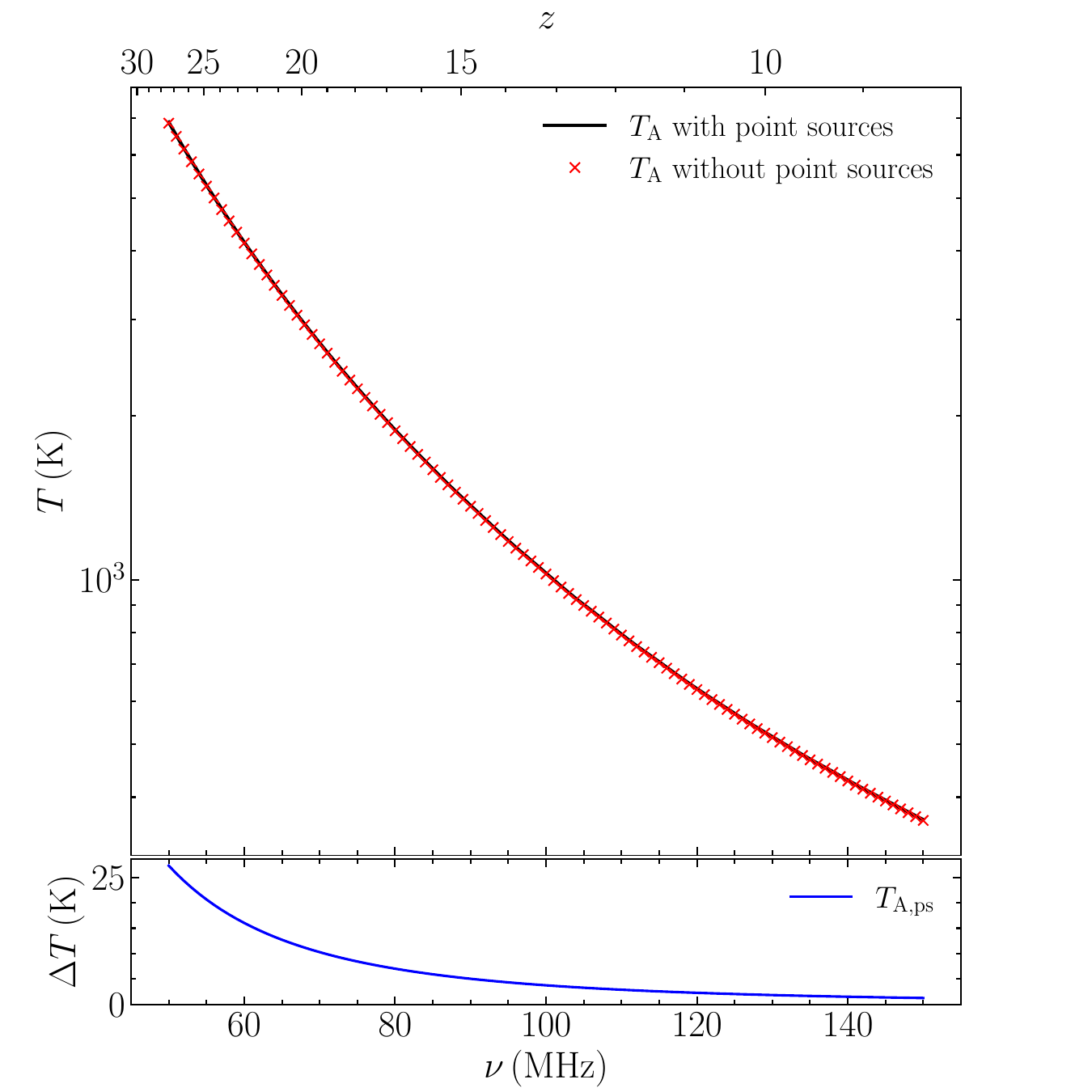}
    \caption{Same as Fig.~\ref{fig:spectrum} but for a hexagonal dipole antenna.}\label{fig:hex1}
\end{figure}

Figure~\ref{fig:hex1} shows the mock antenna temperature with and without the extragalactic point sources contribution. The bottom panel shows the difference between the two data. The minimum, maximum and mean of the difference in $T_{\mathrm{A,ps}}$ for a hexagonal dipole and a conical log-spiral antenna are (approximately) 60, 3 and $\SI{22}{\milli\kelvin}$, respectively. Thus, for the same point sources on the sky an HDA will report a higher antenna temperature than a conical log-spiral antenna throughout the frequency range. 

\begin{figure}
    \centering
    \includegraphics[width=1\linewidth]{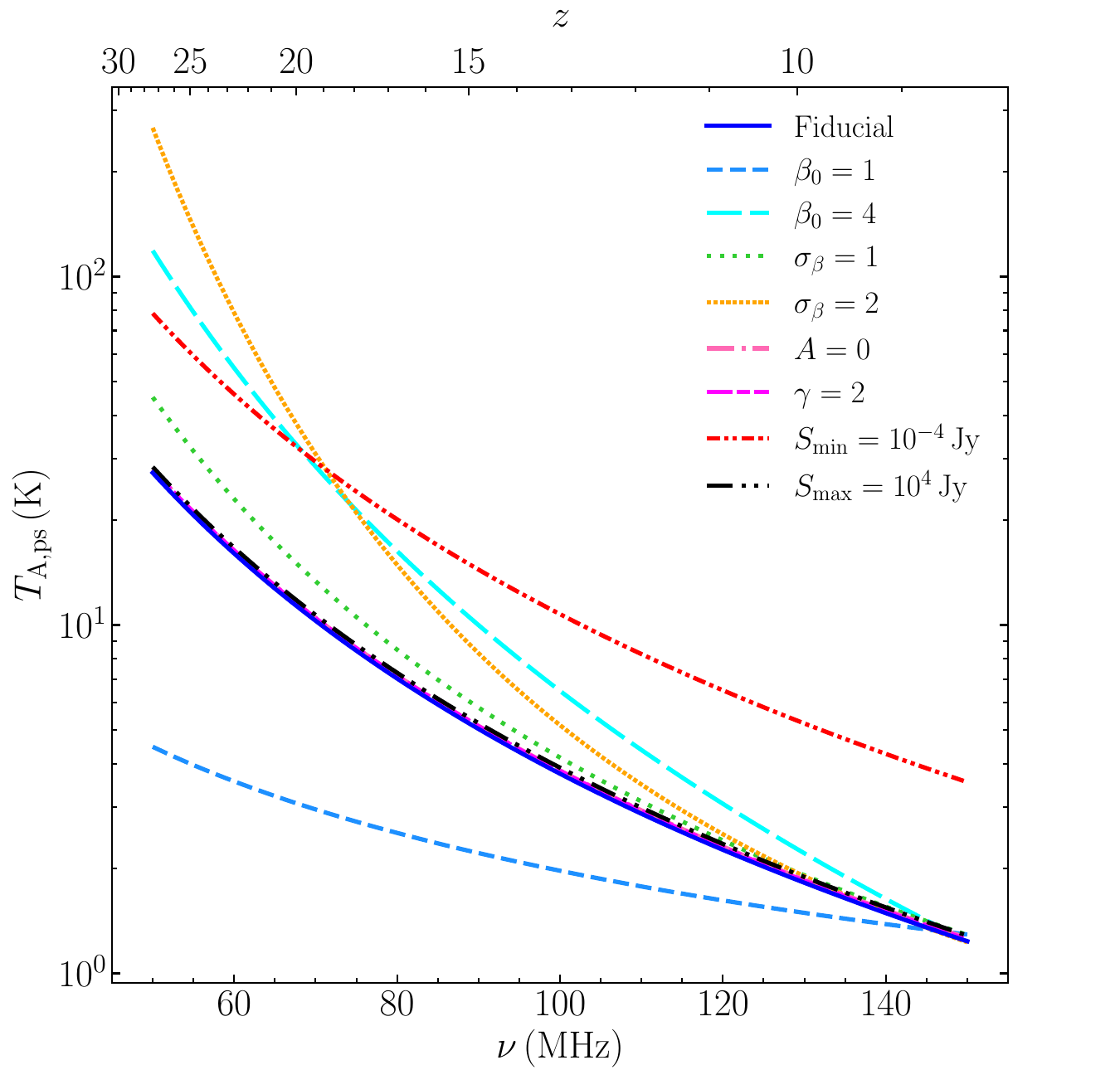}
    \caption{Same as Fig.~\ref{fig:vary_ps} but for a hexagonal dipole antenna. The solid blue curve is repeated from Fig.~\ref{fig:hex1}.}\label{fig:hex2}
\end{figure}

Figure~\ref{fig:hex2} shows antenna temperatures corresponding to a hexagonal dipole antenna for different point sources properties. As was the case with conical log-spiral antenna, all curves conform to the same functional form which is a power-law-with-a-running-index. Table~\ref{tab:vary_ps_hex} shows the result of a least-squares fitting to these curves. The uncertainty on all the numbers is of the order of $10^{-4}$.

\begin{table}
    \caption{Same as Table~\ref{tab:vary_ps} but for a hexagonal dipole antenna. The uncertainty on all the reported numbers is of the order of $10^{-4}$.}\label{tab:vary_ps_hex}
    \centering
    \def\arraystretch{1.1}
    \begin{tabular}{rlll}
        \hline
        Model                      & $T_\mathrm{f}$ & $\beta_\mathrm{f}$ & $\Delta\beta_\mathrm{f}$ \\ \hline
        Fiducial                   & $1.233$        & $2.696$            & $0.114$                  \\
        $\beta_0=1$                & $1.292$        & $0.987$            & $0.129$                  \\
        $\beta_0=4$                & $1.240$        & $4.027$            & $0.114$                  \\
        $\sigma_\beta=1$           & $1.290$        & $2.687$            & $0.499$                  \\
        $\sigma_\beta=2$           & $1.228$        & $2.729$            & $1.976$                  \\
        $A=0$                      & $1.263$        & $2.681$            & $0.125$                  \\
        $\gamma=2$                 & $1.253$        & $2.695$            & $0.118$                  \\
        $S_{\mathrm{min}}=\SI{e-4}{\jansky}$ & $3.548$        & $2.681$            & $0.124$                  \\
        $S_{\mathrm{max}}=\SI{e4}{\jansky}$    & $1.284$        & $2.688$            & $0.119$                  \\
        \hline
    \end{tabular}
\end{table}

\begin{figure}
    \centering
    \includegraphics[width=1\linewidth]{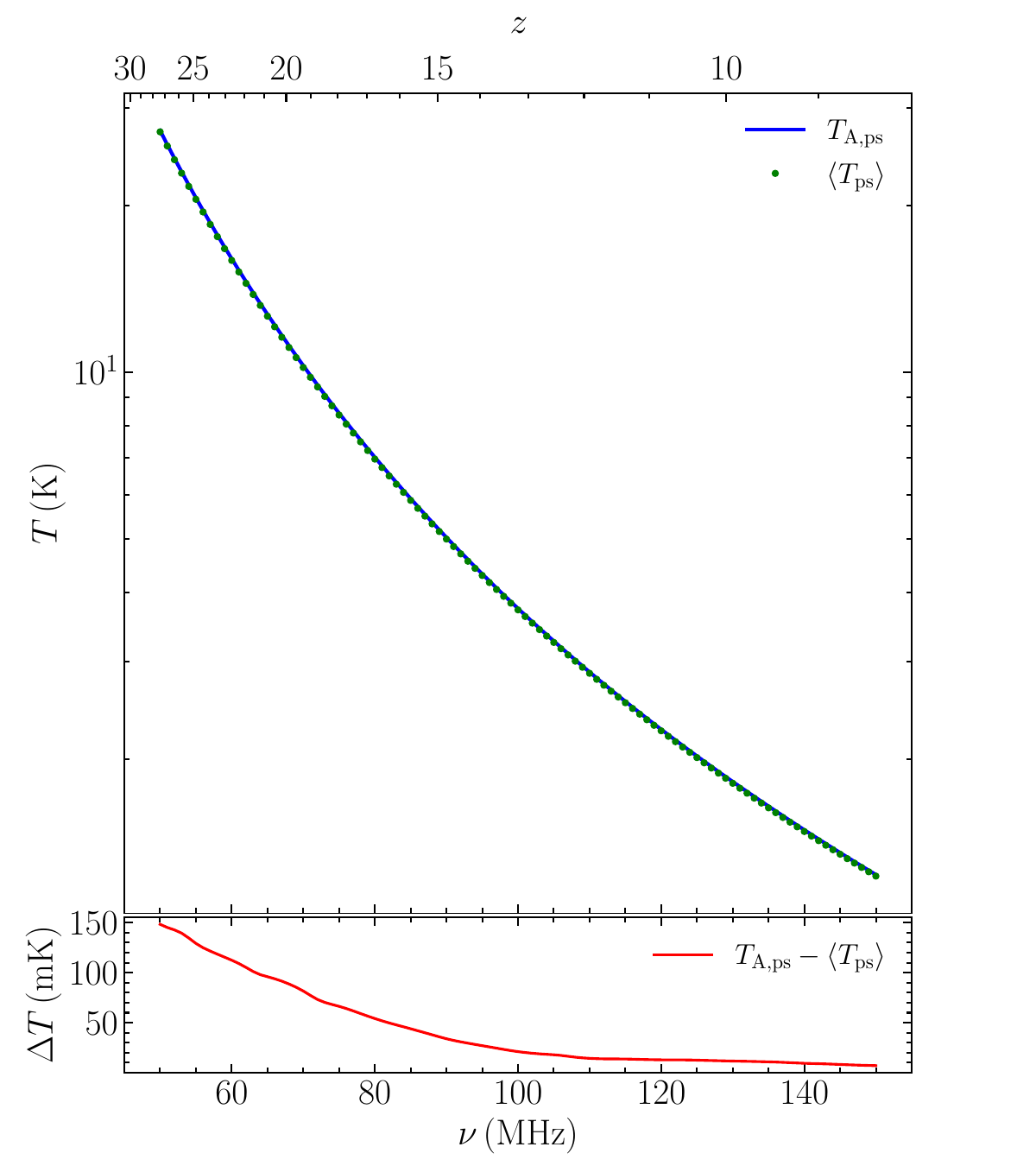}
    \caption{Same as Fig.~\ref{fig:ps} but for a hexagonal dipole antenna. The solid blue curve is repeated from Fig.~\ref{fig:hex1}. Curve with green circles shows the sky average of point sources (repeated from Fig.~\ref{fig:average_ps}).}\label{fig:hex3}
\end{figure}

Figure~\ref{fig:hex3} shows an explicit comparison between averaged sky temperature due to point sources ($\langle T_{\mathrm{ps}}\rangle$) and the corresponding antenna temperature ($T_{\mathrm{A,ps}}$). The maximum chromaticity induced by hexagonal dipole is about $\SI{150}{\milli\kelvin}$ while that for conical log-spiral antenna is about $\SI{100}{\milli\kelvin}$.

\bsp	
\label{lastpage}
\end{document}